\documentclass[manuscript]{acmart}

\usepackage{colortbl}
\usepackage{tabularx}
\usepackage{multirow}
\usepackage{listings}
\usepackage{subcaption}
\usepackage{enumitem}

\usepackage[frozencache,cachedir=.]{minted}

\AtBeginDocument{%
  \providecommand\BibTeX{{%
    \normalfont B\kern-0.5em{\scshape i\kern-0.25em b}\kern-0.8em\TeX}}}

\copyrightyear{2022} 
\acmYear{2022} 
\setcopyright{none}
\acmConference[IUI '22]{27th International Conference on Intelligent User Interfaces}{March 22--25, 2022}{Helsinki, Finland}
\acmBooktitle{27th International Conference on Intelligent User Interfaces (IUI '22), March 22--25, 2022, Helsinki, Finland}
\acmPrice{15.00}
\acmDOI{10.1145/3490099.3511157}
\acmISBN{978-1-4503-9144-3/22/03}

\begin{document}

\title{Better Together? An Evaluation of AI-Supported Code Translation}

\author{Justin D. Weisz}
\orcid{0000-0003-2228-2398}
\affiliation{
    \institution{IBM Research AI}
    \city{Yorktown Heights}
    \state{NY}
    \country{USA}
}
\email{jweisz@us.ibm.com}

\author{Michael Muller}
\orcid{0000-0001-7860-163X}
\affiliation{
    \institution{IBM Research AI}
    \city{Cambridge}
    \state{MA}
    \country{USA}
}
\email{michael_muller@us.ibm.com}

\author{Steven I. Ross}
\orcid{0000-0002-2533-9946}
\affiliation{
    \institution{IBM Research AI}
    \city{Cambridge}
    \state{MA}
    \country{USA}
}
\email{steven_ross@us.ibm.com}

\author{Fernando Martinez}
\orcid{0000-0001-7172-4805}
\affiliation{
    \institution{IBM Argentina}
    \city{La Plata}
    \state{Buenos Aires}
    \country{Argentina}
}
\email{martferc@ar.ibm.com}

\author{Stephanie Houde}
\orcid{0000-0002-0246-2183}
\affiliation{
    \institution{IBM Research AI}
    \city{Cambridge}
    \state{MA}
    \country{USA}
}
\email{Stephanie.Houde@ibm.com}

\author{Mayank Agarwal}
\orcid{0000-0002-8442-2651}
\affiliation{
    \institution{IBM Research AI}
    \city{Cambridge}
    \state{MA}
    \country{USA}
}
\email{Mayank.Agarwal@ibm.com}

\author{Kartik Talamadupula}
\orcid{0000-0002-4628-3785}
\affiliation{
    \institution{IBM Research AI}
    \city{Seattle}
    \state{WA}
    \country{USA}
}
\email{krtalamad@us.ibm.com}

\author{John T. Richards}
\orcid{0000-0001-8489-2170}
\affiliation{
    \institution{IBM Research AI}
    \city{Yorktown Heights}
    \state{NY}
    \country{USA}
}
\email{ajtr@us.ibm.com}

\renewcommand{\shortauthors}{Weisz et al.}

\begin{abstract}
    Generative machine learning models have recently been applied to source code, for use cases including translating code between programming languages, creating documentation from code, and auto-completing methods. Yet, state-of-the-art models often produce code that is erroneous or incomplete. In a controlled study with 32 software engineers, we examined whether such imperfect outputs are helpful in the context of Java-to-Python code translation. When aided by the outputs of a code translation model, participants produced code with fewer errors than when working alone. We also examined how the quality and quantity of AI translations affected the work process and quality of outcomes, and observed that providing multiple translations had a larger impact on the translation process than varying the quality of provided translations. Our results tell a complex, nuanced story about the benefits of generative code models and the challenges software engineers face when working with their outputs. Our work motivates the need for intelligent user interfaces that help software engineers effectively work with generative code models in order to understand and evaluate their outputs and achieve superior outcomes to working alone.
\end{abstract}



\maketitle

\section{Introduction}

Creators of human-centered AI systems aspire to augment the capabilities and performance of their users~\cite{shneiderman2021tutorial, shneiderman2020human, shneiderman2020bridging}. However, how good does an AI model need to be in order to provide such augmentation? Might an imperfect AI system still provide benefit to its users? In this paper, we empirically examine a situation in which human and AI efforts are combined to understand whether the joint effort produces a superior outcome over what either party could accomplish on its own. We examine this notion in the context of software engineering, and specifically in the use case of translating source code from one programming language to another.

Our question is rooted in the observation that state-of-the-art generative code models do not produce 100\% correct output 100\% of the time. For example, the TransCoder model~\cite{roziere2020unsupervised}, which translates code between C++, Java, and Python, only produces a correct translation 30-70\% of the time, depending on which languages it is translating and how many translations it is allowed to produce for a given input. Similarly, a recent evaluation of the GPT-2~\cite{radford2019language}, GPT-3~\cite{brown2020language}, and GPT-Neo~\cite{gao2020pile, gpt-neo} models on their ability to generate code from natural language found that, although these models are becoming increasingly competent at code generation, their overall performance in producing code that passed test cases was low~\cite{hertzberg2010information}. For the foreseeable future, human effort will be needed in order to improve the quality of AI-generated code to the level where it is usable in real systems. Recent work by \citet{weisz2021perfection} asks the question of whether imperfect\footnote{Our view of ``imperfect'' generative models are those that produce outputs that require subsequent human effort to fix. As state-of-the-art generative code models often produce code that contains flaws such as syntax and logical errors, we consider them to be imperfect models. It is beyond the scope of this paper to speculate on whether perfect models are attainable, although our results suggest that such a high standard of quality may not be necessary for them to be useful.} generative models are nonetheless appealing for use by software engineers, and concludes that, \emph{``[c]ounter to expectations, [they] felt a generative model that produced imperfect output [...] \textbf{would} be a useful aid''} (emphasis added). 

While software engineers may be willing to use imperfect models, it is unclear as to whether the use of such models would provide a productivity boost, and how work practices would be affected. In the words of \citet{xu2021ide}, \emph{``There is a surprising paucity of research using human-centered approaches to evaluate the usefulness and impact of [generative] methods within... software development.''} We agree with their assessment, and our work takes the same aim as theirs: examining the extent to which state-of-the-art generative code models impact the performance of software engineers conducting coding tasks. \citeauthor{xu2021ide}'s study examined whether a natural language to code model (NL2Code) helped software engineers complete a variety of programming tasks in Python. Although their experiments were unable to detect any difference in outcomes of completion time, program correctness, and code complexity, we believe further study of  generative code models for other use cases is warranted.

In this paper, we examine professional software engineers using a state-of-the-art model~\cite{roziere2020unsupervised} to conduct a code translation task in which an object-oriented data structure is translated from Java to Python. We compared their work alone to their work when supported by the model. We also examined how the quality and quantity of translations provided impacted their work. Our paper makes the following contributions to the IUI community:

\begin{itemize}
    \item We provide quantitative evidence that \emph{imperfect} generative code models can improve the quality of a software engineer's work in a code translation task, even when the quality and quantity of the model's output varies,
    \item We show how AI support transforms code translation work from being an act of production to an act of review, and identify multiple secondary benefits of AI support beyond the production of code, and 
    \item We identify a need for interactive, intelligent user interfaces that help people work effectively with generative code models in order to achieve superior outcomes.
\end{itemize}

\section{Related Work}
Our work is situated within the emerging area of human-centered AI, which focuses on understanding how AI technologies can augment and enhance human performance and promote human agency \cite{ehsan2020human, riedl2019human, xu2019toward, shneiderman2020bridging, shneiderman2020human, shneiderman2021tutorial, sankaran2020respecting}. Many studies have examined the collaborative relationship between people and AI systems when working on tasks such as decision making or artifact creation. Although these studies often demonstrate how people like using or enjoy working with AI systems, few studies have sought to quantify the extent to which the AI system enhances human performance or the quality of outcomes. Our work fits into this gap by evaluating the effect of AI support on such outcomes: the quality and completeness of a source code translation.

We begin by summarizing recent developments in the ML community that have applied AI technologies to source code. These technologies enable new kinds of AI-supported user experiences for people who work with code. We then summarize human-centered examinations of AI-supported work in two broad categories -- decision making and artifact creation -- and then focus specifically on examinations of AI-supported code work. Our review highlights how the mere presence of AI support does not guarantee superior outcomes, motivating the need to develop a deeper understanding of how to design intelligent user interfaces that help people work effectively with generative AI systems. We highlight several opportunities for enhancing human-AI co-creation in the context of code work in Section~\ref{sec:future-work}.

\subsection{Enabling Technologies for AI-supported Code Work}
\label{sec:ai-for-code}

Within the ML community, much focus has been given recently to applying modern NLP techniques to source code. This idea was promoted by the \emph{naturalness hypothesis}~\cite{devanbu2015new, hindle2016naturalness, allamanis2018survey}, that software is just another form of human communication. Hence, techniques that have been applied to human language ought to work on code as well. Specific techniques, such as automatic machine translation~\cite{nguyen2014migrating, oda2015learning}, have been manifested on code, resulting in the development of models such as TransCoder~\cite{roziere2020unsupervised}, PLBART~\cite{ahmad2021unified}, CodeBERT~\cite{feng2020codebert}, Codex~\cite{chen2021evaluating}, and many others. Models such as these are capable of implementing use cases such as translating code from one programming language to another~\cite{roziere2020unsupervised}, generating documentation for code~\cite{feng2020codebert}, auto-completing code based on a comment or method signature~\cite{chen2021evaluating, kim2021code}, finding duplicated code in a code base~\cite{guo2020graphcodebert}, and generating a set of unit tests~\cite{tufano2020unit}. Although these models offer an impressive set of capabilities, the way in which they are evaluated tends to focus on the accuracy or correctness of their output, rather than examining the extent to which they help people be more effective in their programming work.

\subsection{Human-centered Examinations of AI-supported Tasks}

In defining a research agenda on human-AI collaboration, \citeauthor{seeber2020machines} posed the question, \emph{``What if artificial intelligence (AI) machines became teammates rather than tools?''}~\cite[Abstract]{seeber2020machines}. One important aspect of a successful human-AI partnership is that each party's contributions interact and mesh usefully with the other party's efforts~\cite{parasuraman2008situation}. Many studies have examined the supportive effects of AI systems, with mixed results regarding whether AI augments human task performance or the quality of outcomes. In some studies, the quality of a decision (e.g.~\cite{lai2019human, desmond2021increasing}) or the speed of a process (e.g.~\cite{delarosa2021mixed, wang2021themisto}) were improved with the presence of AI support. In other studies, outcomes were either no different (e.g.~\cite{weber2020draw, xu2021ide}), or worse (e.g.~\cite{clark2018creative}), when people worked with an AI system. We summarize examinations of two different kinds of AI-supported activities: decision making and artifact creation. We also provide a separate summary of work that examines AI-supported code work, even though it can also be characterized as an activity of artifact creation, due to its high relevance to our own study. We provide an overview showing the mixed outcomes of all of this work in Table~\ref{tab:ai-support-studies}.

\begin{table*}[htp]
    \centering
    \begin{tabularx}{\linewidth}{p{1.3cm}p{2.8cm}Xp{2.3cm}}
        \toprule
        \textbf{Domain} & \textbf{Reference} & \textbf{Summary} & \textbf{Effect of AI} \\
        \midrule
        Decision Making & \citet{lai2019human} & An AI model's predictions led to improved performance in a deception detection task & Improved\allowbreak outcomes \\
        \midrule
        Decision Making & \citet{green2019principles} & AI-supported decisions were more accurate, but not more reliable or fair & Mixed outcomes \\
        \midrule
        Decision Making & \citet{desmond2021increasing} \& \citet{ashktorab2021ai} & AI assistance improved speed and accuracy of data labeling tasks & Improved process \& outcomes \\
        \midrule
        Artifact Creation & \citet{delarosa2021mixed} & People created video game levels faster with a generative AI system & Improved process \\
        \midrule
        Artifact Creation & \citet{weber2020draw} & Human-AI teams did not produce higher-quality image restorations over AI-only & Did not improve outcomes \\
        \midrule
        Artifact Creation & \citet{clark2018creative} & Human-AI teams produced less creative marketing slogans over human-only & Reduced outcome quality \\
        \midrule
        Coding & \citet{drosos2020wrex} & Program synthesis tool helped people complete coding tasks faster & Improved process \\
        \midrule
        Coding & \citet{wang2021themisto} & Generative model helped people write documentation faster without loss of quality & Improved process \\
        \midrule
        Coding & \citet{xu2021ide} & Natural language to code model did not improve productivity, speed, or quality & Did not improve process or \allowbreak outcomes \\
        \bottomrule
    \end{tabularx}
    \caption{Selection of studies that examine AI-supported work across decision making, artifact creation, and coding tasks.}
    \label{tab:ai-support-studies}
\end{table*}

\subsubsection{AI-supported Decision Making}
Many AI systems have been developed to aid decision-making processes such as college admissions~\cite{pangburn2019schools}, hiring~\cite{dattner2019hiring}, and mortgage approval~\cite{murawski2019mortgage}, predicated on the beliefs that AI systems make more accurate and less biased judgements than people~\cite{kleinberg2018human, miller2018want, corbett2017even}. Although systems that abdicate \emph{all} decision-making power to AI models have been strongly criticized~\cite{hertzberg2010information, lyn2020risky, kizilaslan2017can, jones2020covid, saxena2020human, saxena2021framework}, studies of whether human-AI joint decision making results in improved decision quality are inconclusive.

Several studies have shown how AI support improves decision-making processes, either by helping people make more accurate decisions, or by helping people make decisions faster. \citet{lai2019human} found that showing people an AI model's predictions improved their performance in a deception detection task. \citet{desmond2021increasing} and \citet{ashktorab2021ai} both found that AI-assistance in data labeling tasks, in which a human annotator made decisions for which labels to apply to data, sped up the data labeling process and increased the accuracy of data labelers.

In contrast, \citet{green2019principles} examined how the interactions between people and AI models influenced the accuracy, reliability, and fairness of their decisions. Although AI-supported decisions were more accurate, people struggled to determine the accuracy of their own decisions as well as the AI's recommendations, \emph{``fail[ing] to calibrate their use of the model to the quality of its predictions... call[ing] into question foundational assumptions about the efficacy and reliability of algorithm-in-the-loop decision making.''}~\cite[][p. 20]{green2019principles}. \citet{wang2021explanations} examined the role of explanations in AI-supported decision making and found that various XAI methods were ineffective in supporting human decision makers on tasks for which they had limited domain expertise. Therefore, it is not always the case that the support provided by an AI model improves human decision-making quality.

\subsubsection{AI-supported Artifact Creation}
Recent advances in generative AI models have led to a massive exploration in how such models can provide assistance in tasks that involve the creation of an artifact. \citet{fan2019collabdraw} and \citet{oh2018lead} examined the use of generative models in drawing and sketching and found that users enjoyed the process of working with the AI system~\cite{oh2018lead} and were able to produce semantically-meaningful sketches~\cite{fan2019collabdraw}. \citet{delarosa2021mixed} and \citet{guzdial2019friend} examined the use of generative models in game level design and found that they helped people more rapidly create a larger number of playable levels~\cite{delarosa2021mixed}, and that the presence of the AI changed the design process~\cite{guzdial2019friend}. Multiple studies of generative AI-supported writing have shown that authors found value in the suggestions of a generative language model~\cite{gero2019metaphoria, gero2019stylistic}.

However, not all human-AI co-creative processes result in quantifiably\allowbreak-improved outcomes. \citet{weber2020draw} examined people working with a generative model to conduct an image restoration task, and found no difference in formal quality metrics between images restored via a generative model alone and those restored via a human-AI team. In the domain of creative writing, \citet{clark2018creative} examined the quality of slogans and stories written with or without the use of a generative model. Although their participants enjoyed the process of working with the model and desired to use it in the future, their evaluations of writing quality with vs. without AI support showed few differences, leading the authors to conclude that \emph{``machine suggestions do not necessarily lead to better written artifacts.''}~\cite[Abstract]{clark2018creative}. In fact, the use of a generative model may have hindered the creative process, as slogans written jointly with the model were rated as being less creative than slogans written by people alone. Thus, we again observe that the presence of AI support does not necessarily guarantee improved quality outcomes.

\subsubsection{AI-supported Code Work}
Recent advances in AI for Code have enabled new kinds of AI assistance in programming work such as writing code and documentation. Wrex~\cite{drosos2020wrex} is a programming-by-example tool that uses program synthesis techniques to generate data science code inside of computational notebooks. In their user study, \citet{drosos2020wrex} found that the tool helped participants complete more tasks, and complete those tasks faster, than when they worked alone. \citet{wang2021themisto} reported similar findings in their evaluation of Themisto, a computational notebook plugin that enables data scientists to generate documentation for code. In that study, participants documented computational notebooks faster when aided by various documentation-generation models, while preserving the same level of quality attained by participants working alone.

In contrast, \citet{kuttal2021trade} conducted a Wizard-of-Oz examination of programmers conducting coding tasks with a (simulated) AI agent, but found no significant improvement to productivity, code quality, or self-efficacy. \citet{xu2021ide} conducted an extensive evaluation of Python programmers who used a natural language to code model to complete a variety of programming tasks. They concluded, \emph{``[w]hile qualitative surveys of developer experience are largely positive, quantitative results with regards to increased productivity, code quality, or program correctness are inconclusive.''}~\cite[Abstract]{xu2021ide}. Hence, we observe a need to further examine situations in which human effort is augmented with AI support to understand the conditions in which superior outcomes are produced.

\section{AI-Supported Code Translation}

As discussed in Section~\ref{sec:ai-for-code}, generative models have been applied to a wide variety of software engineering use cases. In this paper, we specifically focus on AI-assisted code translation. In this use case, code written in one language (e.g. Java) needs to be translated into another language (e.g. Python). Although rule-based methods are commonly used to achieve this functionality, recent work by \citet{roziere2020unsupervised} has shown how unsupervised methods can be used to train transformer models to perform code translation. Neural machine translation (NMT) models trained in this way, coupled with large code datasets (e.g. Project CodeNet~\cite{puri2021project}, AVATAR~\cite{ahmad2021avatar}), are making it easier to create code translation systems across a wide range of languages.

Despite the advances made by NMT, state-of-the-art models do not produce 100\% perfect output 100\% of the time. For example, TransCoder produces correct translations -- those that pass a set of unit tests -- from 30\%-70\% of the time, depending on the source and target languages. In order to arrive at an acceptable correct translation, it is clear that additional human effort is needed to identify and fix the errors contained within an AI translation. This observation leads us to ask, are software engineers more effective when working with AI-produced translations versus translating code themselves? How good does a translation need to be for it to provide value?

Another complication in evaluating the quality of NMT models stems from the fact that these models do not necessary produce a single translation. Due to their use of beam search, these models are able to produce multiple likely translation outputs from a single code input. In evaluating the quality of NMT models, the ML community often uses a \emph{pass@k} metric, in which the model is said to have correctly translated an input if \emph{any} of the top-k outputs passes unit testing. For example, the Codex model~\cite{chen2021evaluating} used by GitHub Copilot\footnote{GitHub Copilot. \url{http://copilot.github.com}} was evaluated with a \emph{pass@k} of 100, and TransCoder was evaluated with a \emph{pass@k} of 25.

Clearly, for a software engineer, reviewing 100, or even 25, translation alternatives is not feasible, especially in the absence of unit tests as is often the case for real-world code translation work. Therefore, we also ask whether multiple translation options is a desirable feature, or if software engineers would be better off focusing their effort on a single, most-likely translation.

Combining all of these observations leads us to ask three research questions about the efficacy of AI support and how it is impacted by the quantity and quality of translations, the impact that AI support has on the work process of translating code, and the perceptions that software engineers have about AI-supported code translation.

\begin{itemize}
    \item \textbf{RQ1}: \emph{Efficacy of AI support}. Are software engineers more effective in translating code from one programming language to another when given access to AI-produced translations? How do the quantity and quality of AI translations affect their work? Are they able to improve upon the quality of imperfect translations?
    \item \textbf{RQ2}: \emph{Impact on work process}. How does the presence of AI translations affect the code translation work process? Does working with AI-produced translations make the translation task easier?
    \item \textbf{RQ3}: \emph{Benefits \& drawbacks}. How do software engineers perceive working with AI-produced code translations? How do they help or hinder their work?
\end{itemize}

\section{AI Support Study}
In order to address our research questions, we conducted a controlled experiment in which participants performed code translation tasks for two data structures within fixed time intervals. We primarily examined the presence of AI support (No AI vs. AI) as a within-subjects factor in order to gauge how AI support shapes the code translation process and impacts the quality of outcomes. We also examined the quantity (1 vs. 5) and quality (worse vs. better) of translations as between-subjects factors in order to understand their role and impact on task performance. Therefore, our study used a mixed factorial deign with 2 within-subjects factors (data structure and the presence of AI support) and 2 between-subjects factors (AI quantity and quality).

\subsection{Code Translation Tasks}
Our code translation task involved the translation of an object-oriented data structure from Java to Python, aided by translations produced by the TransCoder model~\cite{roziere2020unsupervised}. We chose those two languages as they are popular~\cite{cass2021top} and commonly used within our organization, aiding our ability to find participants who had familiarity with both languages. We chose the translation of data structures, as opposed to other kinds of code (e.g. algorithms), for several reasons. First, they do not have specific API dependencies, increasing the likelihood of TransCoder producing workable translations, as well as eliminating the chance that participants' performance was affected by API unfamiliarity. Second, they are comprised of a collection of mostly-independent methods, enabling participants to create a partially-working translation. Finally, data structures are familiar, but not too familiar. Participants are likely to have implemented them as part of their formal computer science education, but they are not likely to have implemented them recently because implementations are already commonly present in modern programming languages or libraries.

Through pretesting, we evaluated several object-oriented data structures for our study and ultimately chose the Trie~\cite{fredkin1960trie, bodon2003trie} and Priority Queue~\cite{cormen2009introduction, ronngren1997comparative} for our translation tasks. These data structures implemented enough differentiated functionality to provide participants with enough work to complete within a study session, yet they were concise enough to not overwhelm participants (or TransCoder). Java implementations of both data structures were written by one author, after studying a number of sample implementations available online. We provide additional details about these data structures and the work required to translate them from Java to Python in Appendix~\ref{appendix:data-structures}, and we provide details on how we employed the TransCoder model~\cite{roziere2020unsupervised} to produce code translations of varied qualities in Appendix~\ref{appendix:ai-produced-translations}.

In order to examine whether participants would be able to achieve superior outcomes when working with AI support, we set a hard limit on how much time participants would have to perform the code translation tasks. Through pretesting, we determined that 30 minutes was long enough to make progress on, but not complete, the translation tasks. By using a short, fixed duration, we avoided observing a ceiling effect in which most participants produced complete and correct translations. Instead, we observed that participants were able to produce more complete translations when working with AI support.

\subsection{Participants}
Code work in our company is conducted by people across many job roles. Thus, we aimed to recruit people who had the requisite skills to complete our code translation task, rather than recruit people within a specific job role. We recruited 182 technologists within our company and screened them using the following criteria to ensure they would be able to perform the code translation task:

\begin{itemize}
    \item More than 1 year of familiarity with both Python and Java, in order to ensure familiarity with both languages.
    \item Have written Java code within the past 5 years, in order to ensure they would be able to read and understand the Java code.
    \item Have written Python code within the past year, in order to ensure they would be able to write Python code.
    \item Have not written their own implementation of the Trie and Priority Queue data structures\footnote{Questions about data structure familiarity were asked in the presence of five other, confounding data structures (e.g. B-Tree, Splay Tree, etc.) to conceal which data structures we would utilize in the study and guard against people refreshing their knowledge before participating in the study.} within the past year, in order to ensure their performance in the study wasn't impacted by recent experience.
    \item Familiarity with the VSCode IDE, in order to remove effects from not being familiar with the code editing environment.
\end{itemize}

Of the 182 people who filled out our screening survey, 55 (30\%) satisfied our selection criteria. From this set, we selected 32 people\footnote{We note that our sample size is comparable (or larger) than other controlled IUI studies (e.g.~\cite{schneider2021increasing, szymanski2021visual, narkar2021model, wiehr2021effect}).} to participate in our study on a first-come, first-served basis. The majority of our participants were from the United States (N=20; 63\%), with smaller numbers from Canada	(N=3; 9\%), Germany (N=2; 6\%), the United Kingdom (N=2; 6\%), and other countries (N=5; 16\%). Twenty-two (69\%) participants identified as male, 9 (28\%) identified as female, and 1 (3\%) preferred not to answer. As all of our participants conducted code work involving the engineering of software as part of their job, we collectively refer to them as ``software engineers.'' However, their specific job roles varied, with 17 (53\%) describing their role as a software engineer, 6 (19\%) as a software architect, 6 (19\%) as a researcher or scientist, and the rest as a developer advocate (1), manager (1), or product manager (1). Eighteen participants (56\%) reported having used AI or machine learning in their work within the past year, and eight participants (25\%) reported having performed code translation work within the past year.

\subsection{Procedure}

\begin{figure*}[htp]
    \centering
    \includegraphics[width=\linewidth]{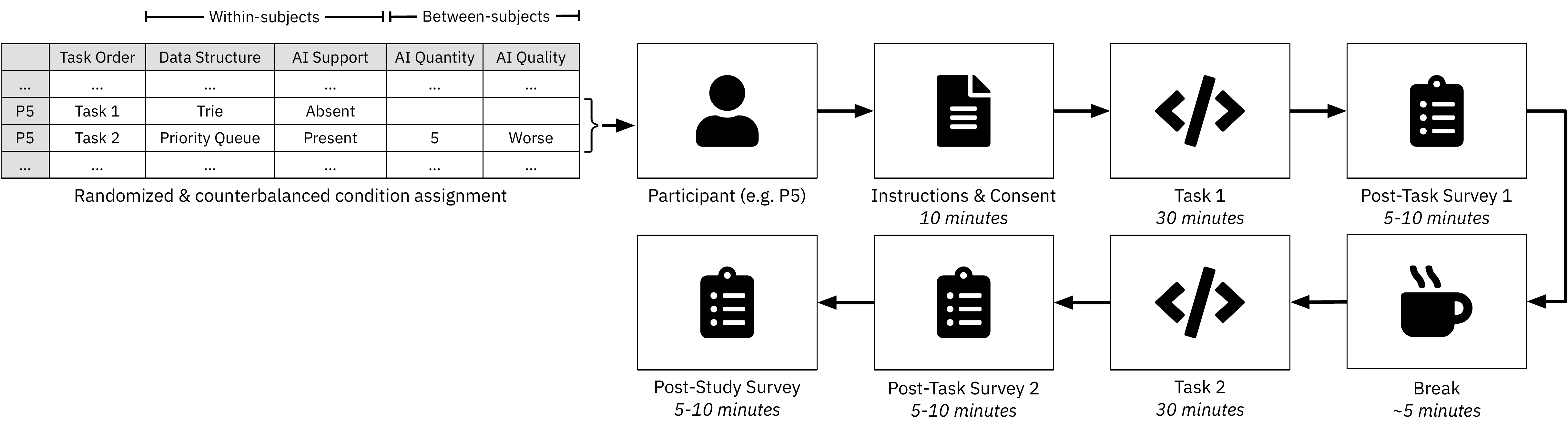}
    \caption{Study Procedure. Our study used a mixed factorial design with 2 within-subjects factors (data structure and the presence of AI support) and 2 between-subjects factors (AI quantity and quality). These factors were randomly assigned to our 32 participants in a counterbalanced fashion, ensuring that equal numbers of participants experienced each combination of factors. In the example above, we show the experience of an an individual participant (P5). The study began with instructions and verbal consent. Next, P5 worked on the first task, translating the Trie without AI support. P5 then filled out a brief survey about the task (Appendix~\ref{appendix:post-task-survey}), followed by a 5-minute break. On the second task, P5 translated the Priority Queue with 5 worse-quality AI translations. After that, they again filled out the post-task survey, which now included additional questions about their experience working with the AI. Finally, they filled out the post-study survey (Appendix~\ref{appendix:post-study-survey}) to contrast their experiences on both tasks.}
    \label{fig:study-design}
\end{figure*}

Participation in our study took place remotely using a video conferencing tool. Each session lasted two hours and was facilitated by one of the authors. The order in which the code translation tasks were completed, as well as the data structure used for each task, was randomized and counterbalanced across participants\footnote{In order to achieve a counterbalanced design, our study required the participant count to be a multiple of 16: 2 task order $\times$ 2 data structure $\times$ 2 AI quantity $\times$ 2 AI quality.}. For the task with AI support, participants were randomly assigned to either have worse or better-quality translations, and to have either 1 or 5 translations. These assignments were also counterbalanced across all participants.

First, we informed participants about the nature of our study by reading a script (listed in Appendix~\ref{appendix:participant-instructions}) that described the expectations of their participation: they would translate two data structures from Java to Python within a fixed 30 minutes, with and without AI support. We twice emphasized the fact that the study was not an evaluation of the participant or their coding skills, but that our aim was to understand the utility of the AI support. Participants provided verbal consent before beginning the study.

In introducing the first translation task, we explained that their translations were to be both complete, by translating all methods and their documentation, and correct, by producing a working implementation. We emphasized the fact that there was more work to do than what could be accomplished within 30 minutes and it was up to the participant to decide how to allocate their time. We also mentioned that they were allowed to use any resources they deemed helpful, such as conducting online searches or running code, but we would not be able to answer any technical questions about the code. We requested participants to think out loud as they conducted their work so that we could understand what they were thinking and how they approached the task.

After these instructions, participants were given a link to an online VSCode editor pre-loaded with the code for their task: in the No AI condition, participants received only the source Java and an empty Python file for their translation in their workspace; in the AI condition, participants received the source Java, one or five AI-produced translations, and an empty Python file for their translation in their workspace. Participants shared their screen so we could observe their work. We started a timer and asked participants to begin, and we provided a 5 minute warning so participants knew when to begin wrapping up their work.

\subsection{Measures}
We recorded a variety of quantitative and qualitative measures to address our research questions about the efficacy of AI support (RQ1), its impact on the work process (RQ2), and peoples' opinions on how it was or was not helpful (RQ3). Some of these measures evaluated the quality of the code translations produced by participants (Section~\ref{sec:code-measures}), and others came from data collected on three surveys: one after each translation task and one at the end of the study (Section~\ref{sec:survey-measures}).

\subsubsection{Code quality measures} 
\label{sec:code-measures}

In order to gauge the quality of all of the code produced in the study -- both by TransCoder and our participants -- we conducted a detailed analysis of the kinds of errors made in the translations. We developed a taxonomy for classifying code translation errors based on the survey of software errors in programming systems provided in \citet[][Table 1]{ko2005framework}. We reviewed all of the software errors they reported, grouped those that were related, and filtered out those that didn't apply to our study (e.g. vendor problems, duplicate tail-digit). Table~\ref{tab:error-taxonomy} lists our error taxonomy with descriptions of how we identified each kind of error.

\begin{table*}[htp]
    \centering
    \begin{tabularx}{\textwidth}{p{2cm}XXp{3cm}}
        \toprule
        \textbf{Error} & \textbf{Description} & \textbf{Operationalization} & \textbf{Source Error(s)} \\
        \midrule
        Translation\allowbreak Error (TE) & Participant mistranslated a code statement by making an error in an assignment statement, a conditional statement, a looping conditional, an array lookup, whitespace, or other logical statement & Count the number of code segments that needed to be modified to fix assignments, conditionals, loops, array lookups, etc. & Translation error, logic error~\cite{panko1998we}; Assignment bug, Iteration bug, Array bug~\cite{gould1975some}; Logical bug~\cite{eisenberg1983apl}; Lexical bugs~\cite{eisenstadt1993tales} \\
        \midrule
        Language Error (LE) & Participant included snippets of Java code within Python or failed to appropriately translate Java language idioms to Pythonic idioms & Count the number of code segments that needed to be modified because Java idioms were used or Python requirements were not met & Dummy bug~\cite{eisenberg1983apl}; Language liability~\cite{knuth1989errors}; Language~\cite{eisenstadt1993tales} \\
        \midrule
        Spurious Error (SE) & Participant included functionality not part of the original Java program (e.g. by defining new methods) & Count the number of irrelevant, unnecessary, or extraneous code statements & Spurious~\cite{johnson1983bug} \\
        \midrule
        Code Omission Error (COE) & Participant omitted the translation of a method or code statements within a method, or provided a trivial implementation (e.g. \texttt{pass}, \texttt{return None}, \texttt{print(``not implemented''}, etc.) & Count the number of instances in which code was added due to missing, trivial, or incomplete method implementations & Missing~\cite{johnson1983bug}; Forgotten function~\cite{knuth1989errors}; Omission error~\cite{panko1998we} \\
        \midrule
        Documentation Omission Error (DOE) & Participant omitted translation of a function’s documentation (e.g. Javadoc comment) & Count the number of Python classes and methods that were missing documentation present in the Java source & Missing~\cite{johnson1983bug}; Omission\allowbreak error~\cite{panko1998we} \\
        \midrule
        Correctness\allowbreak Error (CE) & Participant's translation of a method was incorrect (e.g. did not pass unit tests) & Count the number of methods that required one or more modifications to pass unit tests, including methods that weren't implemented & Algorithm awry~\cite{knuth1989errors} \\
        \bottomrule
    \end{tabularx}
    \caption{Taxonomy of code translation errors. For translation and language errors, we counted code segments rather than lines of code as multiple of these errors could have been present with a single line of code.}
    \label{tab:error-taxonomy}
\end{table*}

In order to identify the errors present in the code translations, we first needed to correct each code artifact in our study (20 baseline AI translations plus 64 participant code translations). We created corrected versions of each code file in two stages: 1) we created a \emph{base} file by copying participants' code, removing any extraneous testing code they may have written, and then appending our own set of unit tests; and 2) we produced a \emph{corrected} file by copying the \emph{base} file, adding in any missing class or method documentation or implementations\footnote{In the case of the Trie, \texttt{delete()} and \texttt{enumerate()} could have been implemented with either one or two methods. We corrected participants' code to preserve the implementation choice they had made. For cases in which participants omitted any implementation, we corrected their code by adding the two-method implementation, as it was closer in spirit to the original Java.}, and then iteratively running it and correcting runtime errors until all unit tests had been passed. When making corrections, we attempted to make the minimal set of changes required to make the code function. By taking the difference between the \emph{base} and \emph{corrected} files, we were then able to identify the locations at which errors occurred, as they represented places in the code that needed to be changed to produce a working translation.

We used GitHub's code review tool\footnote{https://github.com/features/code-review/} to manually classify errors in the code translations. Two authors examined each of the visual \texttt{diffs} produced within GitHub and made comments to label errors and explain the reason for why the code contained that error. Disagreements were discussed and resolved within the comments. We then used GitHub's REST API\footnote{https://docs.github.com/en/rest/} to extract and tally all of the labeled errors. We provide our codebook for the AI-produced translations as part of our reproducibility package (Appendix~\ref{appendix:reproducibility-code}).

We consider two measures of code quality for a code translation. First, we compute the \textbf{overall error rate} of a translation as the total number of errors present divided by its SLOC\footnote{Souce lines of code (SLOC) is a metric of the number of source lines of code; it does not include blank or commented lines. We used the \texttt{cloc} utility to compute SLOC for all code artifacts in our study, available at \url{https://github.com/AlDanial/cloc}.}. Only translation, language, spurious, code and documentation omission errors are included in this metric. The error rate has a minimum of 0 (no errors), but may exceed 1 due to the fact that there may have been more than one error present per source line of code.

\begin{equation*}
    Error\ Rate = \frac{N_{TE} + N_{LE} + N_{SE} + N_{COE} + N_{DOE}}{SLOC}
\end{equation*}

We also consider a secondary measure of translation quality by examining the \textbf{proportion of correctly-translated methods} (PCM), which is computed from the number of correctness errors. Intuitively, PCM is the proportion of methods that passed unit testing without modification.

\begin{equation*}
    PCM = \frac{N_{methods} - N_{CE}}{N_{methods}}
\end{equation*}

We show the baseline measures for all code quality metrics for the AI-produced translations in Table~\ref{tab:ai-translation-errors}. The AI-produced code translations were quite erroneous: the proportion of correctly-translated methods (PCM) varied from 11\% to 67\%, indicating that the AI, even in the best of cases, was unable to complete the code translation task on its own. Error rates ranged between .16 to .60, further indicating the presence of errors that needed to be corrected. As desired for our study design, the error rates for the worse-quality translations (M (SD) = .47 (.07)) were statistically significantly higher than the error rates for the better-quality translations (M (SD) = .27 (.07)), $F(1, 18) = 34.0, p < .001, \eta_{p}^2 = 0.65$ (large).

These error rates represent a baseline for expected human performance when provided with AI support: participants should be able to achieve this level of performance just by adopting AI-produced code without modification. By having these baselines, we are able to address the question of whether participants were able to recognize and remediate the errors present in the code in order to produce an improved translation. We examine this question in Section~\ref{sec:efficacy}.

\begin{table*}[htp]
    \centering
    \begin{tabularx}{\linewidth}{Xlllllllll}
        \toprule
        \textbf{AI Translation} & 
        \multicolumn{3}{l}{\textbf{SLOC}} &
        \multicolumn{3}{l}{\textbf{Error Rate}} &
        \multicolumn{3}{l}{\textbf{PCM}} \\
         & \emph{Count / M} & \emph{SD} & \emph{Range}
         & \emph{Value / M} & \emph{SD} & \emph{Range}
         & \emph{Value / M} & \emph{SD} & \emph{Range} \\
        \midrule
        All translations & 62.6 & 8.3 & [47, 77] & .37 & .12 & [.16, .60] & 47.1\% & 16.9\% & [11.1\%, 66.7\%] \\
        \midrule
        \midrule
        Worse quality  & 63.2 &  5.5 & [57, 72] & .47 & .07 & [.37, .60] & 42.5\% & 18.0\% & [11.1\%, 58.3\%] \\
        Better quality & 62.0 & 10.7 & [47, 77] & .27 & .07 & [.16, .34] & 51.7\% & 15.2\% & [22.2\%, 66.7\%] \\
        \midrule
        \midrule
        \multicolumn{4}{l}{\emph{Trie / Worse}} \\
        1 translation  & 70     & --  & --       & .60 & --   & --         & 58.3\% & --    & -- \\
        5 translations & 68.0   & 2.9 & [65, 72] & .49 & .09  & [.37, .60] & 58.3\% & 0\%   & [58.3\%] \\
        \midrule
        \multicolumn{4}{l}{\emph{Trie / Better}} \\
        1 translation  & 77     & --  & --       & .29 & --   & --         & 66.7\%  & --    & -- \\
        5 translations & 71.0   & 5.2 & [63, 77] & .32 & .02  & [.29, .34] & 63.3\%  & 4.5\% & [58.3\%, 66.7\%] \\
        \midrule
        \multicolumn{4}{l}{\emph{Priority Queue / Worse}} \\
        1 translation  & 58     & --  & --       & .48 & --   & --         & 33\%    & --    & -- \\
        5 translations & 58.4   & 1.1 & [57, 60] & .44 & .04  & [.38, .48] & 26.7\%  & 9.9\% & [11.1\%, 33.3\%] \\
        \midrule
        \multicolumn{4}{l}{\emph{Priority Queue / Better}} \\
        1 translation  & 57     & --  & --       & .16 & --   & --         & 44\%    & --     & -- \\
        5 translations & 53.0   & 5.5 & [47, 57] & .22 & .07  & [.16, .32] & 40.0\%  & 12.7\% & [22.2\%, 55.6\%] \\
        \bottomrule
    \end{tabularx}
    \caption{Baseline error analysis for each set of AI-produced code translations. For the 1 translation condition, SLOC counts and values of error rate and PCM are reported. For all other conditions, means, standard deviations, and ranges are reported.}
    \label{tab:ai-translation-errors}
\end{table*}

\subsubsection{Survey measures}
\label{sec:survey-measures}

After each code translation task, participants filled out the NASA TLX~\cite{hart1988development} scale to report on dimensions of demand, effort, performance, and frustration\footnote{We excluded physical effort as it was not relevant to our task.}. We also asked participants to describe their work process and how they approached the task. For the AI condition, we additionally asked participants to rate the quality of the provided translations and describe how they affected their work. After participants completed both code translation tasks, they filled out a final post-study survey that had them compare their experience in completing the translation task with vs. without AI support. Finally, we collected demographic information such as gender identity, job role, their use of AI or machine learning in their job, and whether they had previously conducted code translation work.

\section{Results}

\subsection{Data \& Analysis}
\label{sec:data-and-analysis}

We collected four types of convergent data from our participants in this study, including 64 code translation artifacts (2 per participant), 64 post-task surveys (2 per participant), 32 post-study surveys, and 32 sets of notes taken by the experimenters during the study sessions.

We use several kinds of statistical models to analyze our quantitative data. In general, when we make comparisons regarding the presence of AI support (No AI vs. AI), we use a linear mixed-effects model that includes participant ID as a random effect to control for within-subject variance. Due to the inclusion of a random effect, degrees of freedom in this model are sometimes fractional and may vary for each factor. When we make comparisons within the AI condition (1 vs. 5 translations or worse vs. better quality), we use a linear model that includes terms for the quality and quantity of AI translations as well as their interaction. In each of these models, we also include terms to control for task order (1st vs. 2nd), data structure (Trie vs. Priority Queue), and their interaction in order to control for learning effects and task characteristics. In reporting statistical results, we report effect sizes as partial eta squared, $\eta_{p}^2$, which is the proportion of variance explained by a given variable after the variance explained by other predictors has been removed. Following \citet{miles2001applying}, we interpret $\eta_{p}^2$ values $\geq .01$ as a small effect, $\geq .06$ as a medium effect, and $\geq .14$ as a large effect. As many of our tests indicated significant, medium-to-large effect sizes, we believe our sample achieved sufficient power.

\subsubsection{Outliers}
During our exploratory data analysis, we identified two significant outliers in the No AI condition: the translations of P9 (0 SLOC) and P23 (5 SLOC). Both of these participants approached the translation task in a methodical way, but struggled to produce a translation when working without AI support and requested to terminate their sessions early. P9 cited a lack of familiarly with Python when they requested to end their session after 17 minutes: \emph{``This would probably take me an hour or more to just look up the comparable Python. I don't use that in my daily work so I'd have to look all this stuff up and I'd rather not even do it... Its just not stuff I have in my head. I'd rather stop.''} P23 expressed a similar sentiment at the 13 minute mark, saying, \emph{``I'm getting nowhere with this and this is frustrating. My Python and my Java skills are simply insufficient to proceed.''} By contrast, when provided with AI support, both participants made significant progress on the task, and actually improved upon the quality of their AI translations. P9 (1, Worse) thoroughly read the translation and was able to find and fix errors in it. P23 (1, Better) launched directly into writing tests, which allowed them to incrementally find and fix errors. These two stories highlight how AI support can provide tremendous value for people who have weaker familiarity with a programming language, even when their general programming skill is quite high; we discuss this theme further in Section~\ref{sec:genai-as-scaffold}. However, due to the fact that their data from the No AI condition are incomparable with that of other participants\footnote{For example, P9's error rate is technically infinite because their code had errors present within a 0 SLOC translation.}, as well as the fact that both participants requested to terminate their No AI sessions early, we exclude those two data points from our statistical analyses. Since P9 and P23 did have successful outcomes when working with the AI, we include their data when making within-AI comparisons\footnote{We also checked if omitting their within-AI data impacted the results of any of our within-AI comparisons and it did not.}.

\subsection{RQ1: Efficacy of AI Support}
\label{sec:efficacy}

Our first research question addresses the efficacy of AI support: are software engineers more effective when they work with the code artifacts produced by a generative code translation model? We operationalized this research question by focusing on our two measures of code translation quality: 1) error rate, the extent to which the code contains the errors described in Table~\ref{tab:error-taxonomy}; and 2) PCM, the proportion of methods that were implemented correctly.

\subsubsection{Effect of AI support on code quality}
\label{sec:effect-aisupport}

Table~\ref{tab:noai-vs-ai} summarizes our main results regarding the effect of AI support on our measures of code quality. When participants had any kind of AI support, the quality of their code was better, as measured by both the error rate and the PCM. Participants' error rates were lower when they had any kind of AI support (M (SD) = .31 (.16)) than when they did not have AI support (M (SD) = .63 (.42)), $F(1, 29.1) = 20.1, p < .001, \eta_{p}^2 = .30$ (large). Participants implemented more correct methods (PCM) when they had any kind of AI support (M (SD) = 50.9\% (16.0\%)) than when they did not have AI support (M (SD) = 34.1\% (29.3\%)), $F(1, 28.7) = 12.1, p = .002, \eta_{p}^2 = .22$ (large). Thus, we observe that the presence of AI support was associated with a 50.8\% reduction in error rate and a 62.1\% increase in PCM.

We also observed that task order played a significant role in participants' code quality. Participants' error rates were higher for their first task (M (SD) = .55 (.42)) than their second task (M (SD) = .38 (.24)), $F(1, 29.2) = 6.12, p = .02, \eta_{p}^2 = .11$ (medium). In addition, participants tended to implement more methods correctly in their second task (M (SD) = 47.2\% (22.5\%)) than their first task (M (SD) = 38.4\% (26.4\%)), $F(1, 28.8) = 3.63, p = .06, \eta_{p}^2 = .08$ (medium). These results indicate the presence of a learning effect in which a participant's experience in the first task helped their work on the second task. Hence, we keep task order as a control term in our models to control for this effect\footnote{As a reminder, our experimental design was a complete factorial. Thus, we are able to make an independent assessment of the effect of task order, and to evaluate other effects, such as AI presence, quantity, and quality, independently of the task order effect.}.

\begin{table*}[htp]
    \centering
    \begin{tabularx}{\linewidth}{Xllllllllllll}
        \toprule
        \textbf{Factor} & \multicolumn{6}{l}{\textbf{Error Rate}} & \multicolumn{6}{l}{\textbf{PCM}} \\
         & $M$ & $SD$ & $df$ & $F$ & $p$ & $\eta_{p}^{2}$ 
         & $M$ & $SD$ & $df$ & $F$ & $p$ & $\eta_{p}^{2}$ \\
        \midrule
        AI presence  & -- & -- & 1, 29.1 & 20.1 & < .001 & .30 & -- & -- & 1, 28.7 & 12.1 & .002 & .22 \\
        \hspace{.2cm} \emph{No AI} & .63 & .42 & & & & & 31.4\% & 29.3\% \\
        \hspace{.2cm} \emph{AI}    & .31 & .16 & & & & & 50.9\% & 16.0\% \\
        \midrule
        Quantity & - & -- & 1, 25 & .006 & n.s. & 0.0 & -- & -- & 1, 25 & .06 & n.s. & .002 \\
        \hspace{.2cm} \emph{1 translation}  & .32 & .19 & & & & & 50.2\% & 16.8\% \\
        \hspace{.2cm} \emph{5 translations} & .50 & .17 & & & & & 51.6\% & 15.7\% \\
        \midrule
        Quality & -- & -- & 1, 25 & 15.8 & .001 & .39 & -- & -- & 1, 25 & 1.4 & n.s. & .05 \\
        \hspace{.3cm} \emph{Worse quality}  & .40 & .16 & & & & & 47.4\% & 17.1\% \\
        \hspace{.3cm} \emph{Better quality} & .22 & .04 & & & & & 54.4\% & 14.6\% \\
        \bottomrule
    \end{tabularx}
    \caption{Summary of participants' code quality metrics across different AI conditions.}
    \label{tab:noai-vs-ai}
\end{table*}

When working with the AI translations, the number of AI-produced translations provided to participants did not have an effect on their error rate, $F(1, 25) = .006, p = n.s., \eta_{p}^{2} = 0.0$. However, the quality of those translations did affect their error rate. Participants with better-quality translations had a lower error rate (M (SD) = .22 (.04)) than participants with worse-quality translations (M (SD) = .40 (.16)), $F(1,25) = 15.8, p = .001, \eta_{p}^2 = .39$ (large). The quantity and quality of AI translations did not have any significant effect on the proportion of methods participants translated correctly (quantity: $F(1, 25) = .06, p = n.s., \eta_{p}^{2} = .002$; quality: $F(1, 25) = 1.4, p = n.s., \eta_{p}^{2} = .05$).

\subsubsection{Individual-level improvements to AI translation quality}

The overall error rates of the AI-produced translations varied from .16 to .60, indicating the potential room for improvement by human effort. Because each participant received a different set of AI translations, we cannot directly compare the baseline error rates of the AI translations as one group against the translations produced when participants had AI support as another group. Instead, we examine whether participants were able to improve upon the quality of \emph{their own set} of AI translations. If a participant's code quality was better than the best-quality AI translation in their set, then they made an \emph{improvement} to code quality. Conversely, if a participant's code quality was worse than the worst-quality AI translation in their set, then they made a \emph{reduction} in code quality. Finally, if a participant's quality fell inside the quality range of their AI translations, then they made \emph{no change} to the code quality. We define this concept mathematically for error rate below; the definition is identical for PCM.

\begin{equation*}
    \small
    ErrorRate_{\Delta} = \left\{
        \begin{array}{lr}
            min_{i \in AI}(ErrorRateAI_{i}) - ErrorRate \\
                \hspace{1cm} \text{if\ } ErrorRate < min_{i \in AI}(ErrorRateAI_{i}) \\
            max_{i \in AI}(ErrorRateAI_{i}) - ErrorRate \\
                \hspace{1cm} \text{if\ } ErrorRate > max_{i \in AI}(ErrorRateAI_{i}) \\
            0\ \text{otherwise} \\
    \end{array}
\right.
\end{equation*}

When $ErrorRate_{\Delta}$ is positive, it indicates that a participant improved code quality, and when it is negative, it indicates that a participant reduced code quality. Overall, a one-sample t test of $ErrorRate_{\Delta}$ indicates that, as a group, participants made a net improvement to the quality of the AI-translated code they received, $t(31) = 2.49, p = .02, 95\%\ CI = [.008, .08]$. A similar test of $PCM_{\Delta}$ does not show a net improvement, $t(31) = -.11, p = n.s., 95\%\ CI = [-.07, .07]$.

\begin{figure*}[htp]
    \centering
    \includegraphics[width=\linewidth]{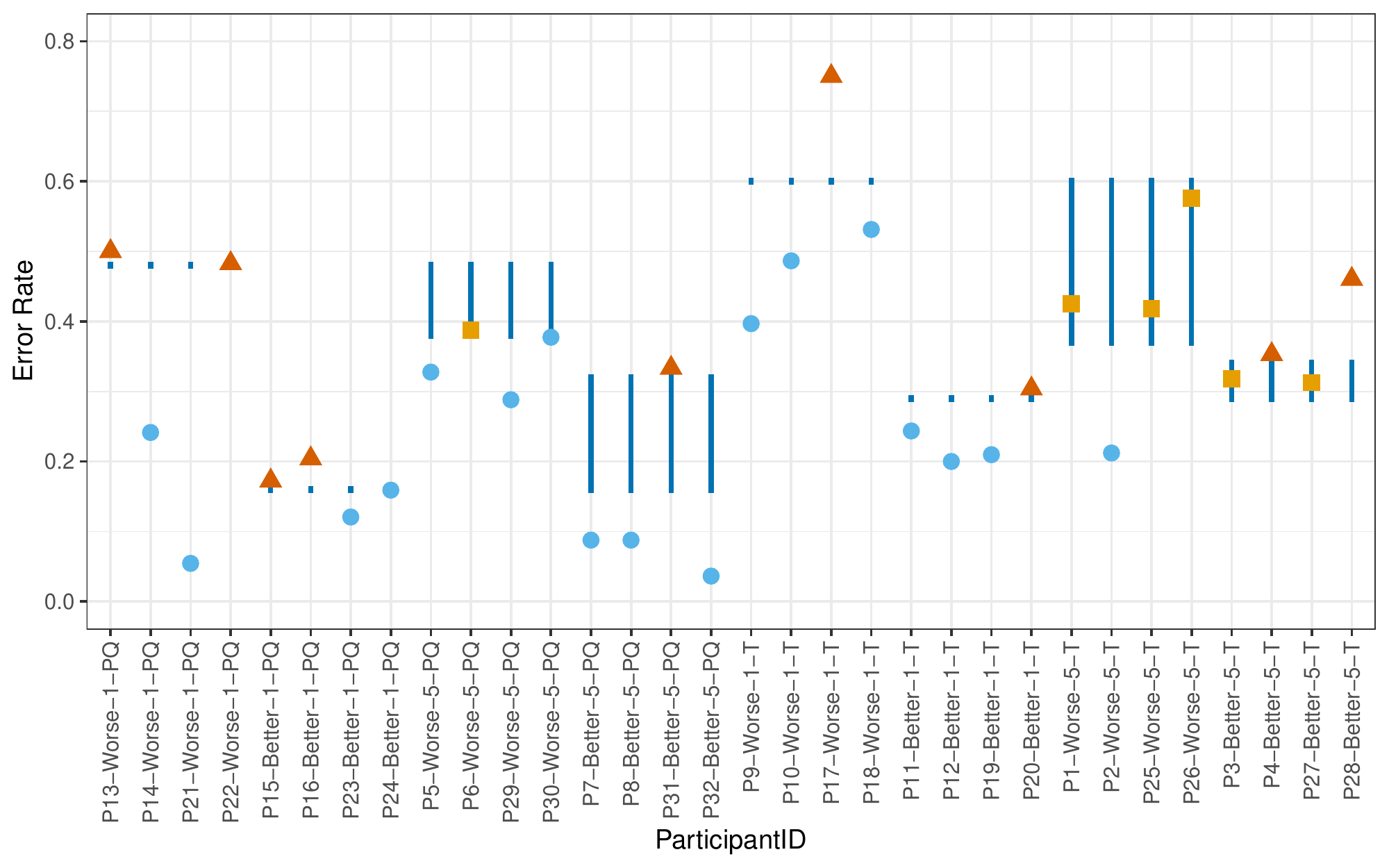}
    \caption{Error rate of participants' code in relation to the error rate(s) of the AI translation(s) they received. Vertical blue lines indicate the range of error rate(s) of each participants' set of AI translations. Points indicate the error rate of each participant's translation. The color and shape of a point indicates whether a participant reduced (red triangles), improved (blue circles), or did not change (orange squares) the quality of their AI translations. Participants are sorted based on the data structure they translated, then by the number of translations they received, and finally by the quality of those translations.}
    \label{fig:expected-error-rate}
\end{figure*}

Figure~\ref{fig:expected-error-rate} shows how each participant's error rate compared with the error rates of their own set of AI translations. Roughly half of participants ($N = 17$, 53\%) made improvements to the quality of their AI translations by reducing the error rate by an average of .11 ($SD = .10$). For 6 participants (19\%), their error rates fell inside the range of their AI translations. For the remaining 9 participants (28\%), their error rates were worse than their set of AI translations, with an average increase in error rate of .04 ($SD = .05$).

What factors might indicate whether a participant was successful in making an improvement in quality? We used a logistic regression model to predict the outcome (improved error rate or not) from the primary factors in our study (AI quality, AI quantity, task order, data structure), as well as participants' self-reported Python experience (recency and frequency of use). This model indicated that recency of Python use was the only significant predictor of a participant's outcome ($\beta = 2.73$, $p = .01$). Participants who had used Python within the past month were 15.3 times more likely to make a quality improvement (by reducing the error rate) than participants who had only used Python within the past year.

We made two observations that may explain why some participants had a quality reduction from their AI translations. One participant explicitly deleted the documentation present in the AI translation (P31, 5, Better), even though our instructions stated that documentation was to be included. Many others simply did not take advantage of the fact that they could run their code, which would have identified many of the errors present within.

\subsubsection{Effect of AI support on kinds of errors}

Working with the AI translations affected the kinds of programming errors made by participants. Recall in Table~\ref{tab:error-taxonomy}, we identified six kinds of errors that could occur in code translation: translation errors (TE), language errors (LE), spurious errors (SE), code (COE) and documentation (DOE) omission errors, and correctness errors (CE). In reviewing all 64 code translations produced by participants, we identified a total of 1,709 errors, and the distribution of these errors differed when participants worked with vs. without AI support, $\chi^2(5, N=1,709) = 244.2, p < .001$.

\begin{table}[htp]
    \centering
    \begin{tabularx}{\linewidth}{Xlllllllll}
        \toprule
        \textbf{Error} & \textbf{No AI} & \textbf{AI} & \boldmath$\chi^2(1)$ & \boldmath$p$ & \textbf{95\% CI} \\
         & $N = 946$ & $N = 763$ &  \\
        \midrule
        Translation Error (TE)   & 11.8\% & 36.8\%  & 148.94 & < .001 & [21.0\%, 29.0\%] \\
        Language Error (LE)      & 35.0\% & 23.7\%  & 25.552 & < .001 & [-15.5\%, -7.0\%] \\
        Spurious Error (SE)      &  2.5\% &  8.4\%  & 29.604 & < .001 & [3.6\%, 8.0\%] \\
        Code Omission Error (COE) & 10.2\% &  2.2\%  & 42.903 & < .001 & [-10.1\%, -5.7\%] \\
        Documentation Omission Error (DOE) & 18.9\% &  7.1\%  & 50.321 & < .001 & [-14.9\%, -8.7\%] \\
        Correctness Error (CE)   & 21.6\% & 21.8\%  &   .009 & n.s.   & [-3.7\%, 4.1\%] \\
        \bottomrule
    \end{tabularx}
    \caption{Distribution of errors observed in code translations produced by participants working with and without AI support. Errors include Translation Error (TE), Language Error (LE), Spurious Error (SE), Code Omission Error (COE), Documentation Omission Error (DOE), and Correctness Error (CE). Two-sample tests for equality of proportions are used to compare the distributions. Magnitudes of the difference between No AI and AI conditions are estimated by 95\% confidence intervals.}
    \label{tab:error-analysis}
\end{table}

In this analysis, we compare between participants working with and without AI support to understand how working with AI-produced translations affected the kinds of errors made. When working without AI support, participants produced a greater number of errors ($N=946$) than when they worked with it ($N=763$); this observation mirrors our earlier finding that participants had significantly lower error rates when working with AI support (.31 vs. .63). However, we now observe clear differences in the \emph{kinds} of errors produced. When working without AI support, a smaller proportion of participants' errors were translation errors (M = 11.8\%) than when working with AI support (M = 36.8\%), suggesting they had an easier time in expressing the intended functionality of the Java code in Python when working alone. They also introduced much less spurious code into their translations when working without AI support (M = 2.5\%) than with (M = 8.4\%). Indeed, one of the difficulties in working with NMT models is that they sometimes produce nonsensical code, which we discuss further in Appendix~\ref{appendix:spurious-code}.

Participants made more language errors when working without the AI (M = 35.0\%) than with (M = 23.7\%), suggesting that they had difficulties in converting Java syntax and idioms to Python. Our own observations confirmed that some participants worked by copying and pasting Java code directly into their Python solutions, then performing the necessary language-level adaptations. However, these language-level adaptations were not always made. For example, we observed cases in which Java's \texttt{null} was used instead of Python's \texttt{None}, Java's \texttt{+=} operator for string append was used instead of Python's \texttt{.append()} method, and Python method bodies were declared using Java's \texttt{\{\}} syntax.

When working without AI support, participants also made more errors of omission. One explanation is that the AI translations contained near-complete, albeit imperfect, translations of each data structure; for the Trie, all methods were defined and documented in each of the 5 AI translations, and for the Priority Queue, all methods were defined but only one method lacked documentation in each of the 5 translations. However, as some participants used a grazing strategy to copy-paste code from across their set of 5 translations, some of that code and documentation was invariably not carried over. Hence, we see that even with AI support, omission errors still occurred.

Our final analysis examines whether participants who improved the quality of the AI translations did so by reducing a specific kind of error. We compared the distribution of errors between the 17 participants who improved the AI's quality and the errors present in the AI translations, and found that there was an overall difference, $\chi^2(5, N=901) = 11.3, p = .04$. However, two-sample tests for equality of proportions only revealed a minor difference in documentation omission errors: participants who improved the AI quality had a higher proportion of documentation omission errors (3.9\%) than the AI translations (1.7\%). Thus, we conclude that there was not a specific kind of error that participants reduced in order to improve the AI's quality, and their improvements were evenly distributed across all error types.

\subsection{RQ2: Impact on Work Process}
We examine the impact of AI support on participants' work practices in several ways. Participants rated their experience through various survey measures, including the NASA TLX~\cite{hart1988development} and questions about how the AI-provided translations affected their work quality and process. We also asked participants to describe their approach to each code translation task, as well as explicitly contrast their experience in working with and without AI support.

\subsubsection{Effect of AI support on the code translation experience}
Mean ratings for each TLX dimension fell roughly in the middle of the 20 point scales (Table~\ref{tab:tlx}), indicating that the code translation task was demanding and required effort, but not to an overwhelming extent. When participants worked with AI support, they rated the task to be as demanding, effortful, and frustrating as when working alone, suggesting that the support provided by the AI did not drastically reduce their own effort.

\begin{table*}[htp]
    \centering
    \begin{tabularx}{\textwidth}{lXllllllll}
        \toprule
        &  & \multicolumn{2}{l}{\textbf{No AI}} & \multicolumn{2}{l}{\textbf{AI}} & \textbf{df} & \textbf{F} & \textbf{p} & \boldmath$\eta_{p}^{2}$ \\
        \textbf{Dimension} & \textbf{Description} & $M$ & $SD$ & $M$ & $SD$ \\
        \midrule
        Mental demand & How mentally demanding was the task? 
          & 12.6 & 4.5 & 12.2 & 4.5 
          & 1,29 & .19 & n.s. & .004 \\
        Temporal demand & How hurried or rushed was the pace of the task? 
          & 12.2 & 5.4 & 11.0 & 4.8 
          & 1,29 & 3.9 & .06  & .11 \\
        Performance & How successful were you in accomplishing what you were asked to do? 
          & 9.9 & 5.8 & 10.5 & 5.1 
          & --  & --  & --   & -- \\
        Effort & How hard did you have to work to accomplish your level of\allowbreak performance? 
          & 12.1 & 3.7 & 10.9 & 4.3 
          & 1,29 & 1.8 & n.s. & .04 \\
        Frustration & How insecure, discouraged, irritated, stressed, and annoyed were you? 
          & 8.6  & 6.1 & 7.0  & 4.4 
          & 1,29 & 1.5 & n.s. & .03 \\
        \bottomrule
    \end{tabularx}
    \caption{Descriptive statistics for NASA TLX~\cite{hart1988development} dimensions of mental demand, temporal demand, performance, effort, and frustration. Physical demand was excluded as it was not applicable to our code translation task. Dimensions were rated on a 20-point scale (1-20), with 20 being the highest-level rating. Linear models were used to compare each TLX dimension and they included terms that controlled for AI presence, task order, data structure, and participant ID as a random effect. The model for performance had a singular fit and is therefore not reported.}
    \label{tab:tlx}
\end{table*}

When working with AI support, participants who were provided with 5 translations reported significantly higher frustration (M (SD) = 9.06 (4.57) of 20) than participants who were provided with one translation (M (SD) = 4.94 (3.26) of 20), $F(1, 26) = 8.59, p < .01, \eta_{p}^2 = .25$ (large). In addition, participants who were provided with 5 translations reported significantly higher mental demand (M (SD) = 14.1 (3.59) of 20) than participants with one translation (M (SD) = 10.4 (4.63) of 20), $F(1, 26) = 5.85, p = .02, \eta_{p}^2 = .18$ (large). Both of these findings indicate that the experience of working with 5 translations was poorer than working with a single one. We did not observe any differences between participants having worse vs. better-quality translations on any TLX measures. These findings were aptly summed up by P7 (5, Better), who explained that, \emph{``[w]ith AI it was faster and efficient... but it also felt that my mind was very stressed to compare all the options I had.''}. P2 (5, Worse) more bluntly rejected the utility of seeing multiple translations: \emph{``the generated code is cool, but [it] isn't too helpful to see 5 different versions of code.''}

From the TLX ratings, it seems that participants did not feel that their performance was better when they worked with AI support, even thought it was. In order to capture a more nuanced understanding of how the presence of AI support affected participants' work, we asked participants to rate whether the AI translations made their work in the task: difficult (-3) vs. easy (+3), slow (-3) vs. fast (+3), and of the worst (-3) vs. of the best (+3) quality. Ratings for all of these dimensions fell on the positive side of the scale (Table~\ref{tab:polar-scales}), indicating that participants generally felt that the AI translations made their work easier, faster, and of a better quality. Many participants talked about how the AI saved them time or helped them work faster. P2 (5, Worse) commented, \textit{``I think with AI support [the translation task is] faster.''} P4 (5, Better) felt, \textit{``it saved me time,''} and P15 (1, Better) stated more strongly that the translation \textit{``[s]aves [me] a lot of time in completing the task, makes the task much easier.''} P12 (1, Better) agreed, offering a reason for why the AI was beneficial: \textit{''[the code] structure [was] already laid out, [I] don't have to think about/look up the [P]ython equivalents of Java classes/etc.''}

\begin{table}[htp]
    \centering
    \begin{tabularx}{\linewidth}{>{\raggedright\hangindent=1em}Xlllll}
        \toprule
        \textbf{Scale} & \textbf{M} & \textbf{SD} & \textbf{t(31)} & \textbf{p} & \textbf{95\% CI} \\
        \midrule
        Difficult~(-3) vs. Easy~(+3)   & 1.34 & 1.60 & 4.75 & < .001 & [0.77, 1.92] \\
        Slow~(-3) vs. Fast~(+3)        & 1.34 & 1.45 & 5.24 & < .001 & [.82, 1.87] \\
        Of the worst~(-3) vs. Of the best~(+3) quality & .72 & 1.22 & 3.32 & .002 & [.28, 1.17] \\
        \bottomrule
    \end{tabularx}
    \caption{Effect of AI translations on ratings of work process and quality. Participants rated the extent to which the AI translations made their work difficult vs. easy, slow vs. fast, and of the worst vs. of the best quality. Mean scores for each dimension fell on the positive ends of each scale, and one-sided t-tests and 95\% confidence intervals confirm that all means were positive and significantly different from 0.}
    \label{tab:polar-scales}
\end{table}

We also asked participants whether they felt they were provided with too many, enough, or not enough translations, and the answer depended on how many translations participants received. Of the 16 participants who received 1 translation, only 3 (19\%) desired to see more. Of the 16 participants who received 5 translations, 9 (56\%) felt that they were provided with too many. Thus, we again observe that the availability of multiple translations was less desirable.

\subsubsection{Effect of AI support on work practices}
\label{sec:work-practices}

Participants with AI support exhibited many different styles and approaches to how they translated the data structures. All participants in the study spent large portions of their time reviewing the AI translations and trying to understand and reason about the AI-produced code. Many participants asked questions to themselves (which we heard due to their thinking aloud) about why the AI had translated code statements in a certain way, especially when they would have translated it differently. We observed two primary styles of working with the AI translations. After reviewing the AI-produced translations, some participants would select a complete translation from which to begin their own work, whereas others were more selective about the AI-produced code they copied into their own translation. In these cases, participants often copy-pasted code from multiple translations in their set. P1 and P3 provide clear examples of these two working styles:

\begin{itemize}
    \item P1 (5, Worse) worked by copying and pasting methods from the AI translations and testing them incrementally as they went. Each time they were ready to paste a new method into their own code, they looked through the available translations and quickly decided which version to copy the next method from. P1 described their work process by saying, \emph{``I'm just trying to be lazy[,] if the AI did a good job then I don't need to do that work.''}
    \item P3 (5, Better) carefully reviewed each AI translation, then chose one as a starting point by copying and pasting it in its entirety into their own code file. They then checked the translation by comparing it with the Java source, starting from the top of the file and moving downward. At each line in the code translation, they cross-checked it with the Java source. When they found an error, they fixed it directly if the solution was clear. In other cases, they reviewed how the translation was done in the other AI translations. Sometimes, they copied code from another translation, which they described as, \emph{``found the better one, I'll steal him.''}
\end{itemize}

The overall sentiment of working with AI support was that it transformed the task from one of translating code to one more akin to code review. P20 offered a straightforward comparison: \emph{``Without AI [it] was like greenfield development, whereas with AI [it] was like fixing legacy buggy code.''} (P20, 1, Better). Working with the AI felt more like a \emph{``proofreading task''} (P12, 1, Better), because \emph{``most of the grunt work was done for me already, all the methods were in place and I just had to review.''} (P18, 1, Worse). Many participants felt that because the AI translation provided a starting point from which to work, it made the task easier.

\begin{quote}
    \emph{``Without AI support I had to understand the Java code... With AI, the biggest benefit was to jump straight to try out the [Python] code and fix errors as they come or as tests fail.''} (P15, 1, Better)
\end{quote}

\begin{quote}
    \emph{``The skeleton of the translation matched quite closely [with] what was provided in the Java implementation, making a method by method comparison trivial.''} (P21, 1, Worse)
\end{quote}

In addition to affecting how they worked, some participants felt that the presence of the AI translations negated the need to even look at the source Java file they were tasked with translating. P28 (5, Better) remarked, \emph{``I found that I was referring to the AI generated code significantly more than the source(java) code when it was available.''} Finally, P18 (1, Worse) identified a trade-off in how the AI's work on the task was helpful, but may not have aided their own understanding of the code: \emph{``AI support was definitely better...[but] from a code understanding... point of view I am not sure if there was a real benefit.''}

\subsection{RQ3: Benefits \& Drawbacks}
\label{sec:benefits-drawbacks}

Our third research question addresses participants' subjective experience in working with the AI translations. Specifically, we examine how they perceived the quality of the AI translations and how those qualities helped or hindered the translation task. 

\subsubsection{Benefits \& utility of AI translations}
Most participants felt that the quality of the AI translations was either acceptable (37.5\%), good (40.6\%), or very good (12.5\%), with only a few participants rating their quality as poor or very poor (9.4\%). To gauge the utility of the AI translations, participants rated the extent to which they felt they were useful, had errors, and helped them complete the task. These ratings were made on a 4-point scale: not at all (1), a little (2), somewhat (3), and a great deal (4). Overall, participants felt the AI translations were useful (a great deal: 50\%, somewhat: 37.5\%, a little: 6.25\%, not at all: 6.25\%), and helped them complete their task (a great deal: 53.1\%, somewhat: 28.1\%, a little: 6.25\%, not at all: 12.5\%), even through they recognized the errors present within them (a great deal: 21.9\%, somewhat: 43.75\%, a little: 28.1\%, not at all: 6.25\%). P19 (1, Better) provided an apt description of the quality of the translations, describing theirs as \emph{``semi-usable.''}

Many participants made positive comments about the utility of the translations, remarking that \emph{``most of the translation code seemed useful''} (P8, 5, Better), \emph{``everything was useful''} (P19, 1, Better), and \emph{``[I] didn't really find anything [to be] not useful.''} (P13, 1, Worse). When asked what the most helpful aspect of the translations were, P28 remarked, \emph{``[the] most useful [aspect] was simply the translation existing.''} (P28, 5, Better).

One specific reason why participants felt the AI translations were useful is because they provided insight into the Python language. Roughly 69\% of participants felt that the AI translations taught them something new about Python (a great deal: 18.7\%, somewhat: 21.9\%, a little: 28.1\%, not at all: 31.3\%). For P5, it was an API: \emph{``I learnt about a new Python API which I didn't know about before.''} (P5, 5, Worse). For P24, the AI translations improved recall of Python-specific syntax: \emph{``The translation remembered some Python syntax I had forgotten.''} (P24, 1, Better). For P7, the AI's ability to produce correct syntax helped them avoid looking it up on their own: \emph{``Syntax finding was easier than Google.''} (P7, 5, Better). P21 appreciated how the AI translation demonstrated the correct syntax for the \emph{``iteration of elements in lists''} (P21, 1, Worse).

Perhaps the greatest benefit of the AI translations was that they provided a \emph{``skeleton''} (P21, 1, Worse), a \emph{``basic structure''} (P1, 5, Worse), a \emph{``starting point''} (P9, 1, Worse), and a \emph{``good start''} (P8, 5, Better). Many participants highlighted how the AI translations reduced the amount of work they needed to do because the \emph{``bulk of the translation was done''} (P13, 1, Worse) and \emph{``[a]ll the boilerplate code was done''} (P25, 5, Worse). P18 (1, Worse) described how the nature of the translation task turned into one of code review, as \emph{``[the] majority of the grunt work was done for me already, all the methods were in place and I just had to review.''} Multiple participants commented on how the AI translations saved them effort because \emph{``[I didn't] hav[e] to type out all the code''} (P29, 5, Worse), they \emph{``saved me some cutting and pasting''} (P23, 1, Better), and they \emph{``saved me typing''} (P10, 1, Worse) because \emph{``some code was already written for me so it saved me the time to write it myself''} (P4, 5, Better). P32 (5, Better) provided an apt summary of this notion: \emph{``It was easier with the AI support because I didn't have to do as much typing.''}

\subsubsection{Drawbacks \& hindrances of AI translations}
Even through most participants felt that it was useful to have the AI translations, and they were helpful in completing the translation task, about half of our participants discussed various problems and difficulties they encountered.

P29 (5, Worse) summarized a trade-off between writing and debugging code when they noted that, \emph{``[l]ess typing involved if a[n] automated translation is available. On the other hands spotting errors in `foreign' code is sometimes challenging.''} Other participants didn't like that the AI translations contained \emph{``[b]ugs, and fairly unreadable code''} (P20, 1, Better) and \emph{``[i]mplementation code that had bugs''} (P1, 5, Worse), resulting in frustration \emph{``[t]hat they didn't work!!!''} (P4, 5, Better).

Other participants expressed similar views, highlighting the fact that finding errors was difficult and time consuming. P25 (5, Worse) said the worst part about the translations was \emph{``[e]rrors in the implementation - and the time spent figuring them out.''} P29 (5, Worse) remarked that it was \emph{``harder to spot error[s] sprinkled in the code.''} P32 (5, Better) didn't like \emph{``[t]hat I had to comb through it to check it didn't have logical errors.''} P21 (1, Worse) noted that the hardest part was in identifying \emph{``small subtle problems in trivial translations''}:

\begin{quote}
    \emph{``There were several cases where it was supposed to translate simple things like `size - 1' but it translated it to `size' or something bogus. For an untrained eye which is skimming over the code those differences may be very hard to spot.''} (P21, 1, Worse)
\end{quote}

This situation caused some participants to doubt their work: \emph{``[t]he incorrect code... made me doubt what I was doing.''} (P3, 5, Better). Some of the specific types of errors produced by the AI also raised doubts about the quality of the model:

\begin{quote}
    \emph{``os.path.join... was a simple bug [that] nearly destroyed my confidence in the tool.''} (P11, 1, Better)
\end{quote}

\begin{quote}
    \emph{``Some of the bugs seemed pretty weird, like not getting the number of arguments correct to a function, or putting the same thing on both sides of a comparator.''} (P14, 1, Worse)
\end{quote}

Participants were also sensitive to the style of the code produced by the AI model. Characteristics such as formatting and the use of Python-specific language features were all discussed as being important for the model to be useful: \emph{``[I]t's least useful when it starts using functions that no [P]ython developers use.''} (P2, 5, Worse); \emph{``[I]t seems to like compressing empty lines which I would find much more useful leaving in.''} (P24, 1, Better); \emph{``The translation [was] not using methods that [the Java class] provided to do what was clearly a 1-to-1 Java to Python translation.''} (P9, 1, Worse).

One final source of frustration regarded the availability of multiple AI translations. Participants who received 5 translations reported higher levels of frustration and mental demand in their NASA TLX scores. P7 (5, Better) clearly articulated the difficulty of working with \emph{``[t]oo many options''} because \emph{``my mind was very stressed to compare all the options I had.''} P5 (5, Worse) highlighted another difficulty, of managing multiple translations across multiple code windows: \emph{``given the many files, scrolling between them was a bit troublesome as one file would get hidden away in the title headers.''} P2 (5, Worse) concurred with this sentiment, and suggested that some sort of \emph{``GUI that can check difference[s] between translations by clicking on code patterns''} would be helpful.

\section{Discussion}

\subsection{Imperfect, but Useful}
Participants were more effective in translating code from Java to Python when aided by AI-produced translations, as evidenced by their code implementing a greater proportion of methods correctly and with fewer errors per SLOC (RQ1). AI support changed the nature of their work process from being one of production to one of review (RQ2), and participants found value in working with the imperfect code translations, even though they recognized their problems (RQ3). In line with how \citet{guzdial2019friend} and \citet{wang2020autoai} both cast AI support systems as actors playing various kinds of supportive roles for their users, we interpret the roles generative code models may play for software engineers in a similar light.

\subsubsection{Generative models as performance amplifiers}
The premise of human-centered AI systems is that they augment or amplify the performance or capabilities of their users~\cite{shneiderman2020bridging, shneiderman2020human, shneiderman2021tutorial}. We observed mixed results in our study. Overall, participants' code translations were of a higher quality when they were provided with translations produced by a generative code model (RQ1). But, which party should get the credit for such quality improvements? The quality of the baseline AI-produced translations (error rate M (SD) = .37 (.12; PCM M (SD) = 47.1\% (16.9\%))) turned out to exceed that of what participants produced alone (error rate M (SD) = .63 (.42); PCM M (SD) = 31.4\% (29.3\%)). Thus, even if a participant made no changes themselves to an AI-produced translation, that translation was likely of a higher quality than what they would have produced when working alone. Therefore, we examined how participants' edits to the AI translations affected their quality, and this examination revealed a more complicated story.

About half of our participants were unable to effectively leverage the head start provided to them by the AI, and yet, the other half were able to improve upon the AI's quality. Plus, for two participants, the support provided by the AI enabled them to complete a task that they were unable to complete on their own and ultimately gave up on. Why were some participants more successful when working with the AI translations than others? Although our analysis suggests that recency of experience is an important factor, we speculate that other factors matter as well. One may have to do with the amount of time participants spent in reviewing the AI translations versus actively working with them. Participants spent much time in reviewing and understanding the AI translations, and given the fixed duration of our study, we observed that some participants did use too much time for this review, and didn't leave themselves enough time for other activities (e.g. writing and running tests, or comparing the code with the Java source). In real-world settings, time is likely to be less of a constraint, and we may be able to uncover additional factors that lead to superior human-AI joint outcomes. Despite not seeing a universal effect, it is clear that imperfect generative code models do have the capability to enhance the performance of their users.

\subsubsection{Generative models as coaches, colleagues, or teachers}
\label{sec:coaches}

Several participants used an analogy of a human role to describe the nature of the support provided by the AI model. P23 (1, Better) said that using the translation was \emph{``like working with a beginning student who had zero context and zero understanding beyond rote knowledge.''} P16 (1, Better) reversed the relationship, suggesting that \emph{``[i]t would be good to have a `coach' type assistant to ask simple questions, and also `observing' how the code should be structured, to help guide newer engineers for proper coding practices.''} P9 (1, Worse) described a more collegial relationship that felt \emph{``almost like pair programming.''} These relational descriptions evoke \citeauthor{lubart2005can}'s proposed roles for computer support as a ``nanny,'' ``coach,'' or ``colleague''~\cite{lubart2005can}, as well as the ``teacher'' role described by \citet{wang2019human}, the ``partner'' role proposed by \citet{main2020guru}, and the ``teammate'' role promoted by \citet{seeber2020machines}. Independent of the specific label used to describe the relationship, many participants felt that they could learn something from the generative code model and the output it produces (RQ3). Indeed, 69\% of participants felt that the AI translations taught them something new they hadn't known before about Python. We believe there is great potential to this idea, that the value of a generative code model may not only lie in its ability to produce code as output, but also in its ability to educate or enlighten its users about the domain for which it was trained. We also observed that the mere presence of the AI translations changed the nature of the task from one of production to one of review (RQ2), akin to how software engineers review each other's work before accepting it. Casting a generative code model as a ``colleague'' or ``teammate'' whose code needs to be reviewed, or as a ``coach'' or a ``teacher'' from which one can learn, may be important in helping people work more effectively with the model.

\subsubsection{Generative models as scaffolds}
\label{sec:genai-as-scaffold}

The cases of P9 and P23 suggest that the support provided by the generative code model functioned as a kind of \emph{scaffold}~\cite{inayatullah2019model, linder2006instructional, warner2017codepilot} that enabled them to complete a task that was otherwise too difficult or daunting (RQ3). Both of these participants were strong software engineers but their Python language skills were not at a level at which they could accomplish the translation task on their own (despite the fact that they met our recruitment criteria). The scaffold provided by the AI translations provided enough support to fill in the gaps in their Python knowledge, enabling them to focus their attention on identifying problems within the AI's translation rather than spending their time refreshing their knowledge of Python. This scaffold even allowed them to make improvements to the quality of their AI translation, demonstrating how a generative code model can reduce knowledge barriers in conducting code work, especially when working with less familiar languages.

\subsection{Limitations}

Our study establishes quantitative evidence that software engineers working with a generative code model were able to produce outcomes that were superior to those achieved when working alone. However, this finding must be understood relative to its context. We examined the simplest form of a code translation user experience -- in essence, ``push button, get results'' -- without any interactive features or means to query the generative model. Yet, such interactive features may be crucially important in helping people be successful when working with the model and its outputs. Many studies have examined how AI models, including generative models, can provide support for their users situated inside an editing environment (e.g., \cite{xu2021ide, wang2021themisto, drosos2020wrex}). Specific to code translation, \citet{weisz2021perfection} suggest that in-IDE features such as confidence highlighting and the selection of translation alternatives could help users, and our own participants requested  other kinds of interactive ways to work with the AI model, such as P1 and P23's desire for an interactive coach that provides \emph{``feedback and hints''} (P1) and P10's desire for a system that interactively provides suggestions on a line-by-line basis. We look forward to future evaluations of how such interactive features impact peoples' experiences and effectiveness in working with generative code models, both for the code translation use case and beyond.

We attempted to control for participants' programming experience to ensure that they were familiar enough with both languages to be able to perform the code translation task. However, as observed in the cases of P9 and P23, peoples' estimations of their own programming experience can vary quite a lot, and measures such as frequency and recency of use may not truly be indicative of one's strength of skill for a given language. We hesitate to conclude that the \emph{only} reason some participants were able to improve the quality of the AI translations was due to the strength of their Python skills, although we were unable to find any other measured factors to explain this result. We encourage future studies to examine a broader range of individual factors that may impact one's ability to have successful joint outcomes when working with a generative code model.

We also attempted to control for the quality of the code produced by the AI model, but given that its quality was generally superior to that of human-only translations -- error rates of AI-produced translations ranged from .16 to .60 compared to a mean No AI error rate of .63 -- we weren't entirely surprised that participants performed better with AI support. However, we believe our study is the first to ask the question of how good a generative model's output needs to be in order for it to be useful, and it seems that even with worse-quality translations, participants were able to make productive use of them. We encourage future studies that explore the tradeoffs between the quality of model output and the quality of joint human-AI outcomes.

\subsection{Future Opportunities}
\label{sec:future-work}

Our study suggests several opportunities for how generative code user experiences can be designed to help people work more effectively, especially via increasing transparency and usability, which are both associated with establishing trust with autonomous tools in the context of high-stakes software engineering~\cite{widder2021trust}. We have also included a reproducibility package in Appendix~\ref{appendix:reproducibility-package} to support further research in AI-supported code translation.

\subsubsection{Intelligent presentation of multiple alternatives}
The inclusion of multiple translation alternatives in our study was associated with increased levels of frustration and mental demand. When designing our study, we speculated that participants might be able to easily distinguish code with vs. without errors by comparing across alternatives to identify areas that were translated differently. Indeed, the way in which many generative code models are evaluated via the $pass@k$ metric presumes that the human's task is one of selection: find the correct output amidst a large set of possibilities. However, the effort required to juggle multiple, similar-looking code translations may have been too high. In addition, the working style of some participants suggested that they preferred to take a more active role in \emph{constructing} a translation rather than \emph{curating} one, by copy-pasting code from multiple translation options. \citet{weisz2021perfection} describe a user experience in which only a single translation is presented in a UI, and highlighting is used to indicate where alternate translations of an individual code segment exist. GitHub Copilot also offers a similar mechanism, using pop-up menus to enable users to browse through autocompletion alternatives. Future work is needed to evaluate how such features impact user experience and effectiveness, although it is clear that some level of intelligent in-IDE support is needed for browsing, comparing, and managing the potentially vast number of outputs of a generative code model.

\subsubsection{Building appropriate trust with a generative model}
All participants examined the translations produced by the generative code model in order to understand what the translations were doing, how they related to the source Java, and whether they represented a correct translation. Much time was spent in performing this evaluation in order to answer the question of, \emph{``[C]an I trust the generated code?''} (P1, 5, Worse). Building trust with automated tools is important for having users adopt those tools into their workflows~\cite{boubin2017quantifying, o2019question}, and prior work shows that providing more information about how an automated system works can improve appropriate trust (e.g.~\cite{drozdal2020trust}). Although the topic of trust is complicated and a detailed examination of it is beyond the scope of this paper, one aspect of trust is predictability: the extent to which people can predict how a system will behave or perform~\cite{cahour2009does}, which comes with experience in using a system. Indeed, this notion was summarized by P1 (5, Worse): \emph{``I would be more comfortable working with AI translated code after having [had] this experience [of participating in this study].''} We believe there is more to be learned about how to calibrate peoples' appropriate trust in a generative code model and we encourage the community to further explore this topic.

\subsubsection{Explainable generative AI}
Many researchers believe that the key to trusting AI systems lies in their ability to explain themselves, their actions, and their decisions~\cite{rossi2018building}. We observed participants asking (of themselves) many of the same kinds of ``why'' and ``how'' questions~\cite{liao2020questioning} that researchers in explainable AI (XAI) are addressing through technological~\cite{arya2020ai}, and even social~\cite{ehsan2021expanding}, means. One recent study investigated what it means for a generative model to be interpretable~\cite{ross2021evaluating}, focusing on whether people are able to interpret how changes in the latent space representation of an artifact manifest changes in its decoded form. Our observation that participants strongly needed to \emph{understand} the model's operation and outputs motivates the need for more formal explorations of explainability for generative models. Work by \citet{liao2020questioning} offers a starting point in enumerating the kinds of questions people have regarding \emph{discriminative} models, and our work motivates the need for further study into the kinds of questions people have regarding \emph{generative} models. For example, participants often wondered about the correspondences between the source Java and the AI's translation, so mechanisms that compute such correspondences for transformer models may serve as one form of explainability.

\subsection{Societal Implications}
Our work has the potential benefits and risks of any technology project that aims to enhance the capabilities of human workers. On the one hand, our results suggest that AI for Code systems can reduce the skills and education necessary to conduct high-quality software translation tasks, thus democratizing this kind of work in the same way that other AI technologies have democratized other tasks in software engineering, machine learning, data science, and even design~\cite{ahmadi2011towards, dibia2018designing, schuler1993participatory, spinuzzi2005methodology, wolf2020democratizing, wang2020autoai}. On the other hand, we realize that the automation of \emph{one part} of a task may cascade into a radical automation of its entirety~\cite{lewney2019technology}, possibly leading to a complete erasure of humans from the task, which would subsequently cause a number of human and societal harms~\cite{bratsberg2021technology, schwabe2020automation, dodel2020perceptions, mutz2021mis, schuler1993participatory}.

To mitigate this risk, we structure our work within a human-centered framework~\cite{shneiderman2020bridging, shneiderman2021tutorial, lynch2021enhance} by asking how generative code models can enhance the effectiveness of software engineers. Consequently, our work \emph{does not} ask the ML community to improve the quality of generative code models; rather, we believe human effort will always be required to at least review, test, and approve a generative code model's output before it is included in a codebase. Therefore, the onus on us, as researchers and designers of human-centered AI systems, is to build and evaluate tools that help our users be maximally effective in working with model outputs that we must assume to be imperfect.


\section{Conclusion}

We conducted a controlled experiment in which 32 professional software engineers translated code from Java to Python with and without the aid of code translations produced by a state-of-the-art neural machine translation model. Participants' code was of a higher quality when they worked with the AI-produced translations, and although the quality of those translations was superior to human-only translations, 59\% of participants made improvements to their quality. Despite the presence of errors in the AI translations, participants overwhelmingly felt that they were useful and helpful in completing the translation task. In addition, two participants who struggled to produce translations on their own were highly successful when working with the AI.

Working with the AI translations shifted the nature of the task from one of writing to reviewing code, although this shift was not associated with a reduction in task demands or the effort required to complete the task. Participants felt that a key benefit of working with the AI was that it provided a starting point for their work, saving them time and effort. However, these perceived savings may not actually have been realized, as participants recognized the time they had to spend in finding and fixing errors contained within the AI's imperfect translations.

Our study demonstrates that the imperfect outputs of a generative code model can provide a performance enhancement to software engineers conducting a code translation task. We have also observed secondary benefits to working with generative code models, such as its capacity to act as a scaffold that enables those with weaker skills to conduct programming tasks, or as a teacher that conveys new knowledge or insights about programming. While the main emphasis of our work is on examining generative code models as performance enhancers, future work is needed to more deeply understand how to design interactive and intelligent user interfaces that maximize the likelihood that joint human-AI effort produces superior outcomes to either party working alone.

\bibliographystyle{ACM-Reference-Format}
\bibliography{references}


\begin{thebibliography}{103}


\ifx \showCODEN    \undefined \def \showCODEN     #1{\unskip}     \fi
\ifx \showDOI      \undefined \def \showDOI       #1{#1}\fi
\ifx \showISBNx    \undefined \def \showISBNx     #1{\unskip}     \fi
\ifx \showISBNxiii \undefined \def \showISBNxiii  #1{\unskip}     \fi
\ifx \showISSN     \undefined \def \showISSN      #1{\unskip}     \fi
\ifx \showLCCN     \undefined \def \showLCCN      #1{\unskip}     \fi
\ifx \shownote     \undefined \def \shownote      #1{#1}          \fi
\ifx \showarticletitle \undefined \def \showarticletitle #1{#1}   \fi
\ifx \showURL      \undefined \def \showURL       {\relax}        \fi
\providecommand\bibfield[2]{#2}
\providecommand\bibinfo[2]{#2}
\providecommand\natexlab[1]{#1}
\providecommand\showeprint[2][]{arXiv:#2}

\bibitem[\protect\citeauthoryear{Ahmad, Chakraborty, Ray, and Chang}{Ahmad
  et~al\mbox{.}}{2021a}]%
        {ahmad2021unified}
\bibfield{author}{\bibinfo{person}{Wasi~Uddin Ahmad}, \bibinfo{person}{Saikat
  Chakraborty}, \bibinfo{person}{Baishakhi Ray}, {and} \bibinfo{person}{Kai-Wei
  Chang}.} \bibinfo{year}{2021}\natexlab{a}.
\newblock \showarticletitle{Unified Pre-training for Program Understanding and
  Generation}.
\newblock \bibinfo{journal}{\emph{arXiv preprint arXiv:2103.06333}}
  (\bibinfo{year}{2021}).
\newblock


\bibitem[\protect\citeauthoryear{Ahmad, Tushar, Chakraborty, and Chang}{Ahmad
  et~al\mbox{.}}{2021b}]%
        {ahmad2021avatar}
\bibfield{author}{\bibinfo{person}{Wasi~Uddin Ahmad},
  \bibinfo{person}{Md~Golam~Rahman Tushar}, \bibinfo{person}{Saikat
  Chakraborty}, {and} \bibinfo{person}{Kai-Wei Chang}.}
  \bibinfo{year}{2021}\natexlab{b}.
\newblock \showarticletitle{AVATAR: A Parallel Corpus for Java-Python Program
  Translation}.
\newblock \bibinfo{journal}{\emph{arXiv preprint arXiv:2108.11590}}
  (\bibinfo{year}{2021}).
\newblock


\bibitem[\protect\citeauthoryear{Ahmadi, Jazayeri, and Repenning}{Ahmadi
  et~al\mbox{.}}{2011}]%
        {ahmadi2011towards}
\bibfield{author}{\bibinfo{person}{Navid Ahmadi}, \bibinfo{person}{Mehdi
  Jazayeri}, {and} \bibinfo{person}{Alexander Repenning}.}
  \bibinfo{year}{2011}\natexlab{}.
\newblock \showarticletitle{Towards democratizing computer science education
  through social game design}. In \bibinfo{booktitle}{\emph{Proceedings of the
  1st International Workshop on Games and Software Engineering}}.
  \bibinfo{pages}{48--51}.
\newblock


\bibitem[\protect\citeauthoryear{Allamanis, Barr, Devanbu, and
  Sutton}{Allamanis et~al\mbox{.}}{2018}]%
        {allamanis2018survey}
\bibfield{author}{\bibinfo{person}{Miltiadis Allamanis},
  \bibinfo{person}{Earl~T Barr}, \bibinfo{person}{Premkumar Devanbu}, {and}
  \bibinfo{person}{Charles Sutton}.} \bibinfo{year}{2018}\natexlab{}.
\newblock \showarticletitle{A survey of machine learning for big code and
  naturalness}.
\newblock \bibinfo{journal}{\emph{ACM Computing Surveys (CSUR)}}
  \bibinfo{volume}{51}, \bibinfo{number}{4} (\bibinfo{year}{2018}),
  \bibinfo{pages}{1--37}.
\newblock


\bibitem[\protect\citeauthoryear{Arya, Bellamy, Chen, Dhurandhar, Hind,
  Hoffman, Houde, Liao, Luss, Mojsilovic, et~al\mbox{.}}{Arya
  et~al\mbox{.}}{2020}]%
        {arya2020ai}
\bibfield{author}{\bibinfo{person}{Vijay Arya}, \bibinfo{person}{Rachel~KE
  Bellamy}, \bibinfo{person}{Pin-Yu Chen}, \bibinfo{person}{Amit Dhurandhar},
  \bibinfo{person}{Michael Hind}, \bibinfo{person}{Samuel~C Hoffman},
  \bibinfo{person}{Stephanie Houde}, \bibinfo{person}{Q~Vera Liao},
  \bibinfo{person}{Ronny Luss}, \bibinfo{person}{Aleksandra Mojsilovic},
  {et~al\mbox{.}}} \bibinfo{year}{2020}\natexlab{}.
\newblock \showarticletitle{AI Explainability 360: An Extensible Toolkit for
  Understanding Data and Machine Learning Models.}
\newblock \bibinfo{journal}{\emph{J. Mach. Learn. Res.}} \bibinfo{volume}{21},
  \bibinfo{number}{130} (\bibinfo{year}{2020}), \bibinfo{pages}{1--6}.
\newblock


\bibitem[\protect\citeauthoryear{Ashktorab, Desmond, Andres, Muller, Joshi,
  Brachman, Sharma, Brimijoin, Pan, Wolf, et~al\mbox{.}}{Ashktorab
  et~al\mbox{.}}{2021}]%
        {ashktorab2021ai}
\bibfield{author}{\bibinfo{person}{Zahra Ashktorab}, \bibinfo{person}{Michael
  Desmond}, \bibinfo{person}{Josh Andres}, \bibinfo{person}{Michael Muller},
  \bibinfo{person}{Narendra~Nath Joshi}, \bibinfo{person}{Michelle Brachman},
  \bibinfo{person}{Aabhas Sharma}, \bibinfo{person}{Kristina Brimijoin},
  \bibinfo{person}{Qian Pan}, \bibinfo{person}{Christine~T Wolf},
  {et~al\mbox{.}}} \bibinfo{year}{2021}\natexlab{}.
\newblock \showarticletitle{AI-Assisted Human Labeling: Batching for Efficiency
  without Overreliance}.
\newblock \bibinfo{journal}{\emph{Proceedings of the ACM on Human-Computer
  Interaction}} \bibinfo{volume}{5}, \bibinfo{number}{CSCW1}
  (\bibinfo{year}{2021}), \bibinfo{pages}{1--27}.
\newblock


\bibitem[\protect\citeauthoryear{Black, Gao, Wang, Leahy, and Biderman}{Black
  et~al\mbox{.}}{2021}]%
        {gpt-neo}
\bibfield{author}{\bibinfo{person}{Sid Black}, \bibinfo{person}{Leo Gao},
  \bibinfo{person}{Phil Wang}, \bibinfo{person}{Connor Leahy}, {and}
  \bibinfo{person}{Stella Biderman}.} \bibinfo{year}{2021}\natexlab{}.
\newblock \bibinfo{booktitle}{\emph{{GPT-Neo: Large Scale Autoregressive
  Language Modeling with Mesh-Tensorflow}}}.
\newblock
\urldef\tempurl%
\url{https://doi.org/10.5281/zenodo.5297715}
\showDOI{\tempurl}
\newblock
\shownote{{If you use this software, please cite it using these metadata.}}


\bibitem[\protect\citeauthoryear{Bodon and R{\'o}nyai}{Bodon and
  R{\'o}nyai}{2003}]%
        {bodon2003trie}
\bibfield{author}{\bibinfo{person}{Ferenc Bodon} {and} \bibinfo{person}{Lajos
  R{\'o}nyai}.} \bibinfo{year}{2003}\natexlab{}.
\newblock \showarticletitle{Trie: an alternative data structure for data mining
  algorithms}.
\newblock \bibinfo{journal}{\emph{Mathematical and Computer Modelling}}
  \bibinfo{volume}{38}, \bibinfo{number}{7-9} (\bibinfo{year}{2003}),
  \bibinfo{pages}{739--751}.
\newblock


\bibitem[\protect\citeauthoryear{Boubin, Rusnock, and Bindewald}{Boubin
  et~al\mbox{.}}{2017}]%
        {boubin2017quantifying}
\bibfield{author}{\bibinfo{person}{Jayson~G Boubin},
  \bibinfo{person}{Christina~F Rusnock}, {and} \bibinfo{person}{Jason~M
  Bindewald}.} \bibinfo{year}{2017}\natexlab{}.
\newblock \showarticletitle{Quantifying compliance and reliance trust behaviors
  to influence trust in human-automation teams}. In
  \bibinfo{booktitle}{\emph{Proceedings of the Human Factors and Ergonomics
  Society Annual Meeting}}, Vol.~\bibinfo{volume}{61}. SAGE Publications Sage
  CA: Los Angeles, CA, \bibinfo{pages}{750--754}.
\newblock


\bibitem[\protect\citeauthoryear{Bratsberg, Rogeberg, and Skirbekk}{Bratsberg
  et~al\mbox{.}}{2021}]%
        {bratsberg2021technology}
\bibfield{author}{\bibinfo{person}{Bernt Bratsberg}, \bibinfo{person}{Ole
  Rogeberg}, {and} \bibinfo{person}{Vegard Skirbekk}.}
  \bibinfo{year}{2021}\natexlab{}.
\newblock \showarticletitle{Technology-induced job loss risk, disability and
  all-cause mortality in Norway}.
\newblock \bibinfo{journal}{\emph{Occupational and Environmental Medicine}}
  (\bibinfo{year}{2021}).
\newblock


\bibitem[\protect\citeauthoryear{Brown, Mann, Ryder, Subbiah, Kaplan, Dhariwal,
  Neelakantan, Shyam, Sastry, Askell, et~al\mbox{.}}{Brown
  et~al\mbox{.}}{2020}]%
        {brown2020language}
\bibfield{author}{\bibinfo{person}{Tom~B Brown}, \bibinfo{person}{Benjamin
  Mann}, \bibinfo{person}{Nick Ryder}, \bibinfo{person}{Melanie Subbiah},
  \bibinfo{person}{Jared Kaplan}, \bibinfo{person}{Prafulla Dhariwal},
  \bibinfo{person}{Arvind Neelakantan}, \bibinfo{person}{Pranav Shyam},
  \bibinfo{person}{Girish Sastry}, \bibinfo{person}{Amanda Askell},
  {et~al\mbox{.}}} \bibinfo{year}{2020}\natexlab{}.
\newblock \showarticletitle{Language models are few-shot learners}.
\newblock \bibinfo{journal}{\emph{arXiv preprint arXiv:2005.14165}}
  (\bibinfo{year}{2020}).
\newblock


\bibitem[\protect\citeauthoryear{Cahour and Forzy}{Cahour and Forzy}{2009}]%
        {cahour2009does}
\bibfield{author}{\bibinfo{person}{B{\'e}atrice Cahour} {and}
  \bibinfo{person}{Jean-Fran{\c{c}}ois Forzy}.}
  \bibinfo{year}{2009}\natexlab{}.
\newblock \showarticletitle{Does projection into use improve trust and
  exploration? An example with a cruise control system}.
\newblock \bibinfo{journal}{\emph{Safety science}} \bibinfo{volume}{47},
  \bibinfo{number}{9} (\bibinfo{year}{2009}), \bibinfo{pages}{1260--1270}.
\newblock


\bibitem[\protect\citeauthoryear{Cass}{Cass}{2021}]%
        {cass2021top}
\bibfield{author}{\bibinfo{person}{Stephen Cass}.}
  \bibinfo{year}{2021}\natexlab{}.
\newblock \showarticletitle{Top Programming Languages 2021}.
\newblock \bibinfo{journal}{\emph{IEEE Spectrum}} (\bibinfo{date}{24 August}
  \bibinfo{year}{2021}).
\newblock
\urldef\tempurl%
\url{https://spectrum.ieee.org/top-programming-languages/}
\showURL{%
\tempurl}


\bibitem[\protect\citeauthoryear{Chen, Tworek, Jun, Yuan, Ponde, Kaplan,
  Edwards, Burda, Joseph, Brockman, Ray, Puri, Krueger, Petrov, Khlaaf, Sastry,
  Mishkin, Chan, Gray, Ryder, Pavlov, Power, Kaiser, Bavarian, Winter, Tillet,
  Such, Cummings, Plappert, Chantzis, Barnes, Herbert-Voss, Guss, Nichol,
  Babuschkin, Balaji, Jain, Carr, Leike, Achiam, Misra, Morikawa, Radford,
  Knight, Brundage, Murati, Mayer, Welinder, McGrew, Amodei, McCandlish,
  Sutskever, and Zaremba}{Chen et~al\mbox{.}}{2021}]%
        {chen2021evaluating}
\bibfield{author}{\bibinfo{person}{Mark Chen}, \bibinfo{person}{Jerry Tworek},
  \bibinfo{person}{Heewoo Jun}, \bibinfo{person}{Qiming Yuan},
  \bibinfo{person}{Henrique Ponde}, \bibinfo{person}{Jared Kaplan},
  \bibinfo{person}{Harri Edwards}, \bibinfo{person}{Yura Burda},
  \bibinfo{person}{Nicholas Joseph}, \bibinfo{person}{Greg Brockman},
  \bibinfo{person}{Alex Ray}, \bibinfo{person}{Raul Puri},
  \bibinfo{person}{Gretchen Krueger}, \bibinfo{person}{Michael Petrov},
  \bibinfo{person}{Heidy Khlaaf}, \bibinfo{person}{Girish Sastry},
  \bibinfo{person}{Pamela Mishkin}, \bibinfo{person}{Brooke Chan},
  \bibinfo{person}{Scott Gray}, \bibinfo{person}{Nick Ryder},
  \bibinfo{person}{Mikhail Pavlov}, \bibinfo{person}{Alethea Power},
  \bibinfo{person}{Lukasz Kaiser}, \bibinfo{person}{Mohammad Bavarian},
  \bibinfo{person}{Clemens Winter}, \bibinfo{person}{Philippe Tillet},
  \bibinfo{person}{Felipe Such}, \bibinfo{person}{Dave Cummings},
  \bibinfo{person}{Matthias Plappert}, \bibinfo{person}{Fotios Chantzis},
  \bibinfo{person}{Elizabeth Barnes}, \bibinfo{person}{Ariel Herbert-Voss},
  \bibinfo{person}{Will Guss}, \bibinfo{person}{Alex Nichol},
  \bibinfo{person}{Igor Babuschkin}, \bibinfo{person}{Suchir Balaji},
  \bibinfo{person}{Shantanu Jain}, \bibinfo{person}{Andrew Carr},
  \bibinfo{person}{Jan Leike}, \bibinfo{person}{Josh Achiam},
  \bibinfo{person}{Vedant Misra}, \bibinfo{person}{Evan Morikawa},
  \bibinfo{person}{Alec Radford}, \bibinfo{person}{Matthew Knight},
  \bibinfo{person}{Miles Brundage}, \bibinfo{person}{Mira Murati},
  \bibinfo{person}{Katie Mayer}, \bibinfo{person}{Peter Welinder},
  \bibinfo{person}{Bob McGrew}, \bibinfo{person}{Dario Amodei},
  \bibinfo{person}{Sam McCandlish}, \bibinfo{person}{Ilya Sutskever}, {and}
  \bibinfo{person}{Wojciech Zaremba}.} \bibinfo{year}{2021}\natexlab{}.
\newblock \showarticletitle{{Evaluating Large Language Models Trained on
  Code}}.
\newblock \bibinfo{journal}{\emph{arXiv preprint arXiv:2107.03374}}
  (\bibinfo{year}{2021}).
\newblock


\bibitem[\protect\citeauthoryear{Clark, Ross, Tan, Ji, and Smith}{Clark
  et~al\mbox{.}}{2018}]%
        {clark2018creative}
\bibfield{author}{\bibinfo{person}{Elizabeth Clark},
  \bibinfo{person}{Anne~Spencer Ross}, \bibinfo{person}{Chenhao Tan},
  \bibinfo{person}{Yangfeng Ji}, {and} \bibinfo{person}{Noah~A Smith}.}
  \bibinfo{year}{2018}\natexlab{}.
\newblock \showarticletitle{Creative writing with a machine in the loop: Case
  studies on slogans and stories}. In \bibinfo{booktitle}{\emph{23rd
  International Conference on Intelligent User Interfaces}}.
  \bibinfo{pages}{329--340}.
\newblock


\bibitem[\protect\citeauthoryear{Corbett-Davies, Goel, and
  Gonz{\'a}lez-Bail{\'o}n}{Corbett-Davies et~al\mbox{.}}{2017}]%
        {corbett2017even}
\bibfield{author}{\bibinfo{person}{Sam Corbett-Davies}, \bibinfo{person}{Sharad
  Goel}, {and} \bibinfo{person}{Sandra Gonz{\'a}lez-Bail{\'o}n}.}
  \bibinfo{year}{2017}\natexlab{}.
\newblock \showarticletitle{Even imperfect algorithms can improve the criminal
  justice system}.
\newblock \bibinfo{journal}{\emph{New York Times}} (\bibinfo{year}{2017}).
\newblock


\bibitem[\protect\citeauthoryear{Cormen, Leiserson, Rivest, and Stein}{Cormen
  et~al\mbox{.}}{2009}]%
        {cormen2009introduction}
\bibfield{author}{\bibinfo{person}{Thomas~H Cormen}, \bibinfo{person}{Charles~E
  Leiserson}, \bibinfo{person}{Ronald~L Rivest}, {and}
  \bibinfo{person}{Clifford Stein}.} \bibinfo{year}{2009}\natexlab{}.
\newblock \bibinfo{booktitle}{\emph{Introduction to algorithms}
  (\bibinfo{edition}{3rd} ed.)}.
\newblock \bibinfo{publisher}{MIT press}, Chapter 6.5: Priority queues.
\newblock


\bibitem[\protect\citeauthoryear{Dattner, Chamorro-Premuzic, Buchband, and
  Schettler}{Dattner et~al\mbox{.}}{2019}]%
        {dattner2019hiring}
\bibfield{author}{\bibinfo{person}{Ben Dattner}, \bibinfo{person}{Tomas
  Chamorro-Premuzic}, \bibinfo{person}{Richard Buchband}, {and}
  \bibinfo{person}{Lucinda Schettler}.} \bibinfo{year}{2019}\natexlab{}.
\newblock \showarticletitle{The Legal and Ethical Implications of Using AI in
  Hiring}.
\newblock \bibinfo{journal}{\emph{Harvard Business Review}} (\bibinfo{date}{25
  April} \bibinfo{year}{2019}).
\newblock
\urldef\tempurl%
\url{https://hbr.org/2019/04/the-legal-and-ethical-implications-of-using-ai-in-hiring}
\showURL{%
Retrieved 05-October-2021 from \tempurl}


\bibitem[\protect\citeauthoryear{Delarosa, Dong, Ruan, Khalifa, and
  Togelius}{Delarosa et~al\mbox{.}}{2021}]%
        {delarosa2021mixed}
\bibfield{author}{\bibinfo{person}{Omar Delarosa}, \bibinfo{person}{Hang Dong},
  \bibinfo{person}{Mindy Ruan}, \bibinfo{person}{Ahmed Khalifa}, {and}
  \bibinfo{person}{Julian Togelius}.} \bibinfo{year}{2021}\natexlab{}.
\newblock \showarticletitle{Mixed-initiative level design with rl brush}. In
  \bibinfo{booktitle}{\emph{International Conference on Computational
  Intelligence in Music, Sound, Art and Design (Part of EvoStar)}}. Springer,
  \bibinfo{pages}{412--426}.
\newblock


\bibitem[\protect\citeauthoryear{Desmond, Muller, Ashktorab, Dugan,
  Duesterwald, Brimijoin, Finegan-Dollak, Brachman, Sharma, Joshi,
  et~al\mbox{.}}{Desmond et~al\mbox{.}}{2021}]%
        {desmond2021increasing}
\bibfield{author}{\bibinfo{person}{Michael Desmond}, \bibinfo{person}{Michael
  Muller}, \bibinfo{person}{Zahra Ashktorab}, \bibinfo{person}{Casey Dugan},
  \bibinfo{person}{Evelyn Duesterwald}, \bibinfo{person}{Kristina Brimijoin},
  \bibinfo{person}{Catherine Finegan-Dollak}, \bibinfo{person}{Michelle
  Brachman}, \bibinfo{person}{Aabhas Sharma}, \bibinfo{person}{Narendra~Nath
  Joshi}, {et~al\mbox{.}}} \bibinfo{year}{2021}\natexlab{}.
\newblock \showarticletitle{Increasing the Speed and Accuracy of Data Labeling
  Through an AI Assisted Interface}. In \bibinfo{booktitle}{\emph{26th
  International Conference on Intelligent User Interfaces}}.
  \bibinfo{pages}{392--401}.
\newblock


\bibitem[\protect\citeauthoryear{Devanbu}{Devanbu}{2015}]%
        {devanbu2015new}
\bibfield{author}{\bibinfo{person}{Premkumar Devanbu}.}
  \bibinfo{year}{2015}\natexlab{}.
\newblock \showarticletitle{New initiative: The naturalness of software}. In
  \bibinfo{booktitle}{\emph{2015 IEEE/ACM 37th IEEE International Conference on
  Software Engineering}}, Vol.~\bibinfo{volume}{2}. IEEE,
  \bibinfo{pages}{543--546}.
\newblock


\bibitem[\protect\citeauthoryear{Dibia, Cox, and Weisz}{Dibia
  et~al\mbox{.}}{2018}]%
        {dibia2018designing}
\bibfield{author}{\bibinfo{person}{Victor Dibia}, \bibinfo{person}{Aaron Cox},
  {and} \bibinfo{person}{Justin Weisz}.} \bibinfo{year}{2018}\natexlab{}.
\newblock \showarticletitle{Designing for Democratization: Introducing Novices
  to Artificial Intelligence Via Maker Kits}.
\newblock \bibinfo{journal}{\emph{arXiv preprint arXiv:1805.10723}}
  (\bibinfo{year}{2018}).
\newblock


\bibitem[\protect\citeauthoryear{Dodel and Mesch}{Dodel and Mesch}{2020}]%
        {dodel2020perceptions}
\bibfield{author}{\bibinfo{person}{Matias Dodel} {and}
  \bibinfo{person}{Gustavo~S Mesch}.} \bibinfo{year}{2020}\natexlab{}.
\newblock \showarticletitle{Perceptions about the impact of automation in the
  workplace}.
\newblock \bibinfo{journal}{\emph{Information, Communication \& Society}}
  \bibinfo{volume}{23}, \bibinfo{number}{5} (\bibinfo{year}{2020}),
  \bibinfo{pages}{665--680}.
\newblock


\bibitem[\protect\citeauthoryear{Drosos, Barik, Guo, DeLine, and
  Gulwani}{Drosos et~al\mbox{.}}{2020}]%
        {drosos2020wrex}
\bibfield{author}{\bibinfo{person}{Ian Drosos}, \bibinfo{person}{Titus Barik},
  \bibinfo{person}{Philip~J Guo}, \bibinfo{person}{Robert DeLine}, {and}
  \bibinfo{person}{Sumit Gulwani}.} \bibinfo{year}{2020}\natexlab{}.
\newblock \showarticletitle{Wrex: A unified programming-by-example interaction
  for synthesizing readable code for data scientists}. In
  \bibinfo{booktitle}{\emph{Proceedings of the 2020 CHI conference on human
  factors in computing systems}}. \bibinfo{pages}{1--12}.
\newblock


\bibitem[\protect\citeauthoryear{Drozdal, Weisz, Wang, Dass, Yao, Zhao, Muller,
  Ju, and Su}{Drozdal et~al\mbox{.}}{2020}]%
        {drozdal2020trust}
\bibfield{author}{\bibinfo{person}{Jaimie Drozdal}, \bibinfo{person}{Justin
  Weisz}, \bibinfo{person}{Dakuo Wang}, \bibinfo{person}{Gaurav Dass},
  \bibinfo{person}{Bingsheng Yao}, \bibinfo{person}{Changruo Zhao},
  \bibinfo{person}{Michael Muller}, \bibinfo{person}{Lin Ju}, {and}
  \bibinfo{person}{Hui Su}.} \bibinfo{year}{2020}\natexlab{}.
\newblock \showarticletitle{Trust in AutoML: exploring information needs for
  establishing trust in automated machine learning systems}. In
  \bibinfo{booktitle}{\emph{Proceedings of the 25th International Conference on
  Intelligent User Interfaces}}. \bibinfo{pages}{297--307}.
\newblock


\bibitem[\protect\citeauthoryear{Ehsan, Liao, Muller, Riedl, and Weisz}{Ehsan
  et~al\mbox{.}}{2021}]%
        {ehsan2021expanding}
\bibfield{author}{\bibinfo{person}{Upol Ehsan}, \bibinfo{person}{Q~Vera Liao},
  \bibinfo{person}{Michael Muller}, \bibinfo{person}{Mark~O Riedl}, {and}
  \bibinfo{person}{Justin~D Weisz}.} \bibinfo{year}{2021}\natexlab{}.
\newblock \showarticletitle{Expanding explainability: Towards social
  transparency in ai systems}. In \bibinfo{booktitle}{\emph{Proceedings of the
  2021 CHI Conference on Human Factors in Computing Systems}}.
  \bibinfo{pages}{1--19}.
\newblock


\bibitem[\protect\citeauthoryear{Ehsan and Riedl}{Ehsan and Riedl}{2020}]%
        {ehsan2020human}
\bibfield{author}{\bibinfo{person}{Upol Ehsan} {and} \bibinfo{person}{Mark~O
  Riedl}.} \bibinfo{year}{2020}\natexlab{}.
\newblock \showarticletitle{Human-centered explainable ai: Towards a reflective
  sociotechnical approach}. In \bibinfo{booktitle}{\emph{International
  Conference on Human-Computer Interaction}}. Springer,
  \bibinfo{pages}{449--466}.
\newblock


\bibitem[\protect\citeauthoryear{Eisenberg and Peelle}{Eisenberg and
  Peelle}{1983}]%
        {eisenberg1983apl}
\bibfield{author}{\bibinfo{person}{Murray Eisenberg} {and}
  \bibinfo{person}{Howard~A Peelle}.} \bibinfo{year}{1983}\natexlab{}.
\newblock \showarticletitle{APL learning bugs}.
\newblock \bibinfo{journal}{\emph{ACM SIGAPL APL Quote Quad}}
  \bibinfo{volume}{13}, \bibinfo{number}{3} (\bibinfo{year}{1983}),
  \bibinfo{pages}{11--16}.
\newblock


\bibitem[\protect\citeauthoryear{Eisenstadt}{Eisenstadt}{1993}]%
        {eisenstadt1993tales}
\bibfield{author}{\bibinfo{person}{Marc Eisenstadt}.}
  \bibinfo{year}{1993}\natexlab{}.
\newblock \showarticletitle{Tales of debugging from the front lines}. In
  \bibinfo{booktitle}{\emph{Empirical Studies of Programmers: Fifth Workshop}}.
  Palo Alto, CA: Ablex Publishing Corporation, \bibinfo{pages}{86--112}.
\newblock


\bibitem[\protect\citeauthoryear{Fan, Dinculescu, and Ha}{Fan
  et~al\mbox{.}}{2019}]%
        {fan2019collabdraw}
\bibfield{author}{\bibinfo{person}{Judith~E Fan}, \bibinfo{person}{Monica
  Dinculescu}, {and} \bibinfo{person}{David Ha}.}
  \bibinfo{year}{2019}\natexlab{}.
\newblock \showarticletitle{Collabdraw: an environment for collaborative
  sketching with an artificial agent}.
\newblock In \bibinfo{booktitle}{\emph{Proceedings of the 2019 on Creativity
  and Cognition}}. \bibinfo{pages}{556--561}.
\newblock


\bibitem[\protect\citeauthoryear{Feng, Guo, Tang, Duan, Feng, Gong, Shou, Qin,
  Liu, Jiang, et~al\mbox{.}}{Feng et~al\mbox{.}}{2020}]%
        {feng2020codebert}
\bibfield{author}{\bibinfo{person}{Zhangyin Feng}, \bibinfo{person}{Daya Guo},
  \bibinfo{person}{Duyu Tang}, \bibinfo{person}{Nan Duan},
  \bibinfo{person}{Xiaocheng Feng}, \bibinfo{person}{Ming Gong},
  \bibinfo{person}{Linjun Shou}, \bibinfo{person}{Bing Qin},
  \bibinfo{person}{Ting Liu}, \bibinfo{person}{Daxin Jiang}, {et~al\mbox{.}}}
  \bibinfo{year}{2020}\natexlab{}.
\newblock \showarticletitle{Codebert: A pre-trained model for programming and
  natural languages}.
\newblock \bibinfo{journal}{\emph{arXiv preprint arXiv:2002.08155}}
  (\bibinfo{year}{2020}).
\newblock


\bibitem[\protect\citeauthoryear{Fredkin}{Fredkin}{1960}]%
        {fredkin1960trie}
\bibfield{author}{\bibinfo{person}{Edward Fredkin}.}
  \bibinfo{year}{1960}\natexlab{}.
\newblock \showarticletitle{Trie memory}.
\newblock \bibinfo{journal}{\emph{Commun. ACM}} \bibinfo{volume}{3},
  \bibinfo{number}{9} (\bibinfo{year}{1960}), \bibinfo{pages}{490--499}.
\newblock


\bibitem[\protect\citeauthoryear{Gao, Biderman, Black, Golding, Hoppe, Foster,
  Phang, He, Thite, Nabeshima, et~al\mbox{.}}{Gao et~al\mbox{.}}{2020}]%
        {gao2020pile}
\bibfield{author}{\bibinfo{person}{Leo Gao}, \bibinfo{person}{Stella Biderman},
  \bibinfo{person}{Sid Black}, \bibinfo{person}{Laurence Golding},
  \bibinfo{person}{Travis Hoppe}, \bibinfo{person}{Charles Foster},
  \bibinfo{person}{Jason Phang}, \bibinfo{person}{Horace He},
  \bibinfo{person}{Anish Thite}, \bibinfo{person}{Noa Nabeshima},
  {et~al\mbox{.}}} \bibinfo{year}{2020}\natexlab{}.
\newblock \showarticletitle{The Pile: An 800GB Dataset of Diverse Text for
  Language Modeling}.
\newblock \bibinfo{journal}{\emph{arXiv preprint arXiv:2101.00027}}
  (\bibinfo{year}{2020}).
\newblock


\bibitem[\protect\citeauthoryear{Gero and Chilton}{Gero and Chilton}{2019a}]%
        {gero2019stylistic}
\bibfield{author}{\bibinfo{person}{Katy~Ilonka Gero} {and}
  \bibinfo{person}{Lydia~B Chilton}.} \bibinfo{year}{2019}\natexlab{a}.
\newblock \showarticletitle{How a Stylistic, Machine-Generated Thesaurus
  Impacts a Writer's Process}.
\newblock In \bibinfo{booktitle}{\emph{Proceedings of the 2019 on Creativity
  and Cognition}}. \bibinfo{pages}{597--603}.
\newblock


\bibitem[\protect\citeauthoryear{Gero and Chilton}{Gero and Chilton}{2019b}]%
        {gero2019metaphoria}
\bibfield{author}{\bibinfo{person}{Katy~Ilonka Gero} {and}
  \bibinfo{person}{Lydia~B Chilton}.} \bibinfo{year}{2019}\natexlab{b}.
\newblock \showarticletitle{Metaphoria: An algorithmic companion for metaphor
  creation}. In \bibinfo{booktitle}{\emph{Proceedings of the 2019 CHI
  Conference on Human Factors in Computing Systems}}. \bibinfo{pages}{1--12}.
\newblock


\bibitem[\protect\citeauthoryear{Gould}{Gould}{1975}]%
        {gould1975some}
\bibfield{author}{\bibinfo{person}{John~D Gould}.}
  \bibinfo{year}{1975}\natexlab{}.
\newblock \showarticletitle{Some psychological evidence on how people debug
  computer programs}.
\newblock \bibinfo{journal}{\emph{International Journal of Man-Machine
  Studies}} \bibinfo{volume}{7}, \bibinfo{number}{2} (\bibinfo{year}{1975}),
  \bibinfo{pages}{151--182}.
\newblock


\bibitem[\protect\citeauthoryear{Green and Chen}{Green and Chen}{2019}]%
        {green2019principles}
\bibfield{author}{\bibinfo{person}{Ben Green} {and} \bibinfo{person}{Yiling
  Chen}.} \bibinfo{year}{2019}\natexlab{}.
\newblock \showarticletitle{The principles and limits of algorithm-in-the-loop
  decision making}.
\newblock \bibinfo{journal}{\emph{Proceedings of the ACM on Human-Computer
  Interaction}} \bibinfo{volume}{3}, \bibinfo{number}{CSCW}
  (\bibinfo{year}{2019}), \bibinfo{pages}{1--24}.
\newblock


\bibitem[\protect\citeauthoryear{Guo, Ren, Lu, Feng, Tang, Liu, Zhou, Duan,
  Svyatkovskiy, Fu, et~al\mbox{.}}{Guo et~al\mbox{.}}{2020}]%
        {guo2020graphcodebert}
\bibfield{author}{\bibinfo{person}{Daya Guo}, \bibinfo{person}{Shuo Ren},
  \bibinfo{person}{Shuai Lu}, \bibinfo{person}{Zhangyin Feng},
  \bibinfo{person}{Duyu Tang}, \bibinfo{person}{Shujie Liu},
  \bibinfo{person}{Long Zhou}, \bibinfo{person}{Nan Duan},
  \bibinfo{person}{Alexey Svyatkovskiy}, \bibinfo{person}{Shengyu Fu},
  {et~al\mbox{.}}} \bibinfo{year}{2020}\natexlab{}.
\newblock \showarticletitle{Graphcodebert: Pre-training code representations
  with data flow}.
\newblock \bibinfo{journal}{\emph{arXiv preprint arXiv:2009.08366}}
  (\bibinfo{year}{2020}).
\newblock


\bibitem[\protect\citeauthoryear{Guzdial, Liao, Chen, Chen, Shah, Shah, Reno,
  Smith, and Riedl}{Guzdial et~al\mbox{.}}{2019}]%
        {guzdial2019friend}
\bibfield{author}{\bibinfo{person}{Matthew Guzdial}, \bibinfo{person}{Nicholas
  Liao}, \bibinfo{person}{Jonathan Chen}, \bibinfo{person}{Shao-Yu Chen},
  \bibinfo{person}{Shukan Shah}, \bibinfo{person}{Vishwa Shah},
  \bibinfo{person}{Joshua Reno}, \bibinfo{person}{Gillian Smith}, {and}
  \bibinfo{person}{Mark~O Riedl}.} \bibinfo{year}{2019}\natexlab{}.
\newblock \showarticletitle{Friend, collaborator, student, manager: How design
  of an ai-driven game level editor affects creators}. In
  \bibinfo{booktitle}{\emph{Proceedings of the 2019 CHI conference on human
  factors in computing systems}}. \bibinfo{pages}{1--13}.
\newblock


\bibitem[\protect\citeauthoryear{Hart and Staveland}{Hart and
  Staveland}{1988}]%
        {hart1988development}
\bibfield{author}{\bibinfo{person}{Sandra~G Hart} {and}
  \bibinfo{person}{Lowell~E Staveland}.} \bibinfo{year}{1988}\natexlab{}.
\newblock \showarticletitle{Development of NASA-TLX (Task Load Index): Results
  of empirical and theoretical research}.
\newblock In \bibinfo{booktitle}{\emph{Advances in psychology}}.
  Vol.~\bibinfo{volume}{52}. \bibinfo{publisher}{Elsevier},
  \bibinfo{pages}{139--183}.
\newblock


\bibitem[\protect\citeauthoryear{Hertzberg, Liberti, and Paravisini}{Hertzberg
  et~al\mbox{.}}{2010}]%
        {hertzberg2010information}
\bibfield{author}{\bibinfo{person}{Andrew Hertzberg},
  \bibinfo{person}{Jose~Maria Liberti}, {and} \bibinfo{person}{Daniel
  Paravisini}.} \bibinfo{year}{2010}\natexlab{}.
\newblock \showarticletitle{Information and incentives inside the firm:
  Evidence from loan officer rotation}.
\newblock \bibinfo{journal}{\emph{The Journal of Finance}}
  \bibinfo{volume}{65}, \bibinfo{number}{3} (\bibinfo{year}{2010}),
  \bibinfo{pages}{795--828}.
\newblock


\bibitem[\protect\citeauthoryear{Hindle, Barr, Gabel, Su, and Devanbu}{Hindle
  et~al\mbox{.}}{2016}]%
        {hindle2016naturalness}
\bibfield{author}{\bibinfo{person}{Abram Hindle}, \bibinfo{person}{Earl~T
  Barr}, \bibinfo{person}{Mark Gabel}, \bibinfo{person}{Zhendong Su}, {and}
  \bibinfo{person}{Premkumar Devanbu}.} \bibinfo{year}{2016}\natexlab{}.
\newblock \showarticletitle{On the naturalness of software}.
\newblock \bibinfo{journal}{\emph{Commun. ACM}} \bibinfo{volume}{59},
  \bibinfo{number}{5} (\bibinfo{year}{2016}), \bibinfo{pages}{122--131}.
\newblock


\bibitem[\protect\citeauthoryear{Inayatullah, Azam, and Anwar}{Inayatullah
  et~al\mbox{.}}{2019}]%
        {inayatullah2019model}
\bibfield{author}{\bibinfo{person}{Mohammad Inayatullah},
  \bibinfo{person}{Farooque Azam}, {and} \bibinfo{person}{Muhammad~Waseem
  Anwar}.} \bibinfo{year}{2019}\natexlab{}.
\newblock \showarticletitle{Model-based scaffolding code generation for
  cross-platform applications}. In \bibinfo{booktitle}{\emph{2019 IEEE 10th
  Annual Information Technology, Electronics and Mobile Communication
  Conference (IEMCON)}}. IEEE, \bibinfo{pages}{1006--1012}.
\newblock


\bibitem[\protect\citeauthoryear{Johnson, Soloway, Cutler, and Draper}{Johnson
  et~al\mbox{.}}{1983}]%
        {johnson1983bug}
\bibfield{author}{\bibinfo{person}{W~Lewis Johnson}, \bibinfo{person}{Elliot
  Soloway}, \bibinfo{person}{Benjamin Cutler}, {and} \bibinfo{person}{Steven
  Draper}.} \bibinfo{year}{1983}\natexlab{}.
\newblock \bibinfo{booktitle}{\emph{Bug catalogue: I}}.
\newblock \bibinfo{publisher}{Yale University Press}.
\newblock


\bibitem[\protect\citeauthoryear{Jones, King, Baker, and Ingham}{Jones
  et~al\mbox{.}}{2020}]%
        {jones2020covid}
\bibfield{author}{\bibinfo{person}{Bernadette Jones},
  \bibinfo{person}{Paula~Toko King}, \bibinfo{person}{Gabrielle Baker}, {and}
  \bibinfo{person}{Tristram Ingham}.} \bibinfo{year}{2020}\natexlab{}.
\newblock \showarticletitle{COVID-19, Intersectionality, and Health Equity for
  Indigenous Peoples with Lived Experience of Disability}.
\newblock \bibinfo{journal}{\emph{American Indian Culture and Research
  Journal}} \bibinfo{volume}{44}, \bibinfo{number}{2} (\bibinfo{year}{2020}),
  \bibinfo{pages}{71--88}.
\newblock


\bibitem[\protect\citeauthoryear{Kim, Zhao, Tian, and Chandra}{Kim
  et~al\mbox{.}}{2021}]%
        {kim2021code}
\bibfield{author}{\bibinfo{person}{Seohyun Kim}, \bibinfo{person}{Jinman Zhao},
  \bibinfo{person}{Yuchi Tian}, {and} \bibinfo{person}{Satish Chandra}.}
  \bibinfo{year}{2021}\natexlab{}.
\newblock \showarticletitle{Code prediction by feeding trees to transformers}.
  In \bibinfo{booktitle}{\emph{2021 IEEE/ACM 43rd International Conference on
  Software Engineering (ICSE)}}. IEEE, \bibinfo{pages}{150--162}.
\newblock


\bibitem[\protect\citeauthoryear{Kizilaslan and Lookman}{Kizilaslan and
  Lookman}{2017}]%
        {kizilaslan2017can}
\bibfield{author}{\bibinfo{person}{Atay Kizilaslan} {and}
  \bibinfo{person}{Aziz~A Lookman}.} \bibinfo{year}{2017}\natexlab{}.
\newblock \showarticletitle{Can Economically Intuitive Factors Improve Ability
  of Proprietary Algorithms to Predict Defaults of Peer-to-Peer Loans?}
\newblock \bibinfo{journal}{\emph{Available at SSRN 2987613}}
  (\bibinfo{year}{2017}).
\newblock


\bibitem[\protect\citeauthoryear{Kleinberg, Lakkaraju, Leskovec, Ludwig, and
  Mullainathan}{Kleinberg et~al\mbox{.}}{2018}]%
        {kleinberg2018human}
\bibfield{author}{\bibinfo{person}{Jon Kleinberg}, \bibinfo{person}{Himabindu
  Lakkaraju}, \bibinfo{person}{Jure Leskovec}, \bibinfo{person}{Jens Ludwig},
  {and} \bibinfo{person}{Sendhil Mullainathan}.}
  \bibinfo{year}{2018}\natexlab{}.
\newblock \showarticletitle{Human decisions and machine predictions}.
\newblock \bibinfo{journal}{\emph{The quarterly journal of economics}}
  \bibinfo{volume}{133}, \bibinfo{number}{1} (\bibinfo{year}{2018}),
  \bibinfo{pages}{237--293}.
\newblock


\bibitem[\protect\citeauthoryear{Knuth}{Knuth}{1989}]%
        {knuth1989errors}
\bibfield{author}{\bibinfo{person}{Donald~E Knuth}.}
  \bibinfo{year}{1989}\natexlab{}.
\newblock \showarticletitle{The errors of TEX}.
\newblock \bibinfo{journal}{\emph{Software: Practice and Experience}}
  \bibinfo{volume}{19}, \bibinfo{number}{7} (\bibinfo{year}{1989}),
  \bibinfo{pages}{607--685}.
\newblock


\bibitem[\protect\citeauthoryear{Ko and Myers}{Ko and Myers}{2005}]%
        {ko2005framework}
\bibfield{author}{\bibinfo{person}{Andrew~J Ko} {and} \bibinfo{person}{Brad~A
  Myers}.} \bibinfo{year}{2005}\natexlab{}.
\newblock \showarticletitle{A framework and methodology for studying the causes
  of software errors in programming systems}.
\newblock \bibinfo{journal}{\emph{Journal of Visual Languages \& Computing}}
  \bibinfo{volume}{16}, \bibinfo{number}{1-2} (\bibinfo{year}{2005}),
  \bibinfo{pages}{41--84}.
\newblock


\bibitem[\protect\citeauthoryear{Kuttal, Ong, Kwasny, and Robe}{Kuttal
  et~al\mbox{.}}{2021}]%
        {kuttal2021trade}
\bibfield{author}{\bibinfo{person}{Sandeep~Kaur Kuttal}, \bibinfo{person}{Bali
  Ong}, \bibinfo{person}{Kate Kwasny}, {and} \bibinfo{person}{Peter Robe}.}
  \bibinfo{year}{2021}\natexlab{}.
\newblock \showarticletitle{Trade-offs for Substituting a Human with an Agent
  in a Pair Programming Context: The Good, the Bad, and the Ugly}. In
  \bibinfo{booktitle}{\emph{Proceedings of the 2021 CHI Conference on Human
  Factors in Computing Systems}}. \bibinfo{pages}{1--20}.
\newblock


\bibitem[\protect\citeauthoryear{Lai and Tan}{Lai and Tan}{2019}]%
        {lai2019human}
\bibfield{author}{\bibinfo{person}{Vivian Lai} {and} \bibinfo{person}{Chenhao
  Tan}.} \bibinfo{year}{2019}\natexlab{}.
\newblock \showarticletitle{On human predictions with explanations and
  predictions of machine learning models: A case study on deception detection}.
  In \bibinfo{booktitle}{\emph{Proceedings of the conference on fairness,
  accountability, and transparency}}. \bibinfo{pages}{29--38}.
\newblock


\bibitem[\protect\citeauthoryear{Lewney, Alexandri, and Storrie}{Lewney
  et~al\mbox{.}}{2019}]%
        {lewney2019technology}
\bibfield{author}{\bibinfo{person}{Richard Lewney}, \bibinfo{person}{Eva
  Alexandri}, {and} \bibinfo{person}{Donald~W Storrie}.}
  \bibinfo{year}{2019}\natexlab{}.
\newblock \bibinfo{booktitle}{\emph{Technology Scenario: Employment
  Implications of Radical Automation}}.
\newblock \bibinfo{publisher}{Publications Office of the European Union}.
\newblock


\bibitem[\protect\citeauthoryear{Liao, Gruen, and Miller}{Liao
  et~al\mbox{.}}{2020}]%
        {liao2020questioning}
\bibfield{author}{\bibinfo{person}{Q~Vera Liao}, \bibinfo{person}{Daniel
  Gruen}, {and} \bibinfo{person}{Sarah Miller}.}
  \bibinfo{year}{2020}\natexlab{}.
\newblock \showarticletitle{Questioning the AI: informing design practices for
  explainable AI user experiences}. In \bibinfo{booktitle}{\emph{Proceedings of
  the 2020 CHI Conference on Human Factors in Computing Systems}}.
  \bibinfo{pages}{1--15}.
\newblock


\bibitem[\protect\citeauthoryear{Linder, Abbott, and Fromberger}{Linder
  et~al\mbox{.}}{2006}]%
        {linder2006instructional}
\bibfield{author}{\bibinfo{person}{Stephen~Paul Linder}, \bibinfo{person}{David
  Abbott}, {and} \bibinfo{person}{Michael~J Fromberger}.}
  \bibinfo{year}{2006}\natexlab{}.
\newblock \showarticletitle{An instructional scaffolding approach to teaching
  software design}.
\newblock \bibinfo{journal}{\emph{Journal of Computing Sciences in Colleges}}
  \bibinfo{volume}{21}, \bibinfo{number}{6} (\bibinfo{year}{2006}),
  \bibinfo{pages}{238--250}.
\newblock


\bibitem[\protect\citeauthoryear{Lubart}{Lubart}{2005}]%
        {lubart2005can}
\bibfield{author}{\bibinfo{person}{Todd Lubart}.}
  \bibinfo{year}{2005}\natexlab{}.
\newblock \showarticletitle{How can computers be partners in the creative
  process: classification and commentary on the special issue}.
\newblock \bibinfo{journal}{\emph{International Journal of Human-Computer
  Studies}} \bibinfo{volume}{63}, \bibinfo{number}{4-5} (\bibinfo{year}{2005}),
  \bibinfo{pages}{365--369}.
\newblock


\bibitem[\protect\citeauthoryear{Lyn}{Lyn}{2020}]%
        {lyn2020risky}
\bibfield{author}{\bibinfo{person}{Alexandra Lyn}.}
  \bibinfo{year}{2020}\natexlab{}.
\newblock \showarticletitle{Risky Business: Artificial Intelligence and Risk
  Assessments in Sentencing and Bail Procedures in the United States}.
\newblock \bibinfo{journal}{\emph{Available at SSRN 3831441}}
  (\bibinfo{year}{2020}).
\newblock


\bibitem[\protect\citeauthoryear{Lynch}{Lynch}{2021}]%
        {lynch2021enhance}
\bibfield{author}{\bibinfo{person}{Shana Lynch}.}
  \bibinfo{year}{2021}\natexlab{}.
\newblock \bibinfo{booktitle}{\emph{Enhance, not Replace: AI’s Potential to
  Make Our Work – and Lives – Better}}.
\newblock
\urldef\tempurl%
\url{https://hai.stanford.edu/news/enhance-not-replace-ais-potential-make-our-work-and-lives-better}
\showURL{%
Retrieved October 2, 2021 from \tempurl}


\bibitem[\protect\citeauthoryear{Main and Grierson}{Main and Grierson}{2020}]%
        {main2020guru}
\bibfield{author}{\bibinfo{person}{Angus Main} {and} \bibinfo{person}{Mick
  Grierson}.} \bibinfo{year}{2020}\natexlab{}.
\newblock \showarticletitle{Guru, Partner, or Pencil Sharpener? Understanding
  Designers' Attitudes Towards Intelligent Creativity Support Tools}.
\newblock \bibinfo{journal}{\emph{arXiv preprint arXiv:2007.04848}}
  (\bibinfo{year}{2020}).
\newblock


\bibitem[\protect\citeauthoryear{Miles and Shevlin}{Miles and Shevlin}{2001}]%
        {miles2001applying}
\bibfield{author}{\bibinfo{person}{Jeremy Miles} {and} \bibinfo{person}{Mark
  Shevlin}.} \bibinfo{year}{2001}\natexlab{}.
\newblock \bibinfo{booktitle}{\emph{Applying regression and correlation: A
  guide for students and researchers}}.
\newblock \bibinfo{publisher}{Sage}.
\newblock


\bibitem[\protect\citeauthoryear{Miller}{Miller}{2018}]%
        {miller2018want}
\bibfield{author}{\bibinfo{person}{Alex~P Miller}.}
  \bibinfo{year}{2018}\natexlab{}.
\newblock \showarticletitle{Want less-biased decisions? Use algorithms}.
\newblock \bibinfo{journal}{\emph{Harvard business review}}
  \bibinfo{volume}{26} (\bibinfo{year}{2018}).
\newblock


\bibitem[\protect\citeauthoryear{Murawski}{Murawski}{2019}]%
        {murawski2019mortgage}
\bibfield{author}{\bibinfo{person}{John Murawski}.}
  \bibinfo{year}{2019}\natexlab{}.
\newblock \showarticletitle{Mortgage Providers Look to AI to Process Home Loans
  Faster}.
\newblock \bibinfo{journal}{\emph{Wall Street Journal}} (\bibinfo{date}{18
  March} \bibinfo{year}{2019}).
\newblock
\urldef\tempurl%
\url{https://www.wsj.com/articles/mortgage-providers-look-to-ai-to-process-home-loans-faster-11552899212}
\showURL{%
Retrieved 05-October-2021 from \tempurl}


\bibitem[\protect\citeauthoryear{Mutz}{Mutz}{2021}]%
        {mutz2021mis}
\bibfield{author}{\bibinfo{person}{Diana~C Mutz}.}
  \bibinfo{year}{2021}\natexlab{}.
\newblock \showarticletitle{(Mis) Attributing the Causes of American Job Loss:
  The Consequences of Getting It Wrong}.
\newblock \bibinfo{journal}{\emph{Public Opinion Quarterly}}
  \bibinfo{volume}{85}, \bibinfo{number}{1} (\bibinfo{year}{2021}),
  \bibinfo{pages}{101--122}.
\newblock


\bibitem[\protect\citeauthoryear{Narkar, Zhang, Liao, Wang, and Weisz}{Narkar
  et~al\mbox{.}}{2021}]%
        {narkar2021model}
\bibfield{author}{\bibinfo{person}{Shweta Narkar}, \bibinfo{person}{Yunfeng
  Zhang}, \bibinfo{person}{Q~Vera Liao}, \bibinfo{person}{Dakuo Wang}, {and}
  \bibinfo{person}{Justin~D Weisz}.} \bibinfo{year}{2021}\natexlab{}.
\newblock \showarticletitle{Model LineUpper: Supporting Interactive Model
  Comparison at Multiple Levels for AutoML}. In \bibinfo{booktitle}{\emph{26th
  International Conference on Intelligent User Interfaces}}.
  \bibinfo{pages}{170--174}.
\newblock


\bibitem[\protect\citeauthoryear{Nguyen, Nguyen, and Nguyen}{Nguyen
  et~al\mbox{.}}{2014}]%
        {nguyen2014migrating}
\bibfield{author}{\bibinfo{person}{Anh~Tuan Nguyen},
  \bibinfo{person}{Tung~Thanh Nguyen}, {and} \bibinfo{person}{Tien~N Nguyen}.}
  \bibinfo{year}{2014}\natexlab{}.
\newblock \showarticletitle{Migrating code with statistical machine
  translation}. In \bibinfo{booktitle}{\emph{Companion Proceedings of the 36th
  International Conference on Software Engineering}}.
  \bibinfo{pages}{544--547}.
\newblock


\bibitem[\protect\citeauthoryear{Oda, Fudaba, Neubig, Hata, Sakti, Toda, and
  Nakamura}{Oda et~al\mbox{.}}{2015}]%
        {oda2015learning}
\bibfield{author}{\bibinfo{person}{Yusuke Oda}, \bibinfo{person}{Hiroyuki
  Fudaba}, \bibinfo{person}{Graham Neubig}, \bibinfo{person}{Hideaki Hata},
  \bibinfo{person}{Sakriani Sakti}, \bibinfo{person}{Tomoki Toda}, {and}
  \bibinfo{person}{Satoshi Nakamura}.} \bibinfo{year}{2015}\natexlab{}.
\newblock \showarticletitle{Learning to generate pseudo-code from source code
  using statistical machine translation}. In \bibinfo{booktitle}{\emph{2015
  30th IEEE/ACM International Conference on Automated Software Engineering
  (ASE)}}. IEEE, \bibinfo{pages}{574--584}.
\newblock


\bibitem[\protect\citeauthoryear{Oh, Song, Choi, Kim, Lee, and Suh}{Oh
  et~al\mbox{.}}{2018}]%
        {oh2018lead}
\bibfield{author}{\bibinfo{person}{Changhoon Oh}, \bibinfo{person}{Jungwoo
  Song}, \bibinfo{person}{Jinhan Choi}, \bibinfo{person}{Seonghyeon Kim},
  \bibinfo{person}{Sungwoo Lee}, {and} \bibinfo{person}{Bongwon Suh}.}
  \bibinfo{year}{2018}\natexlab{}.
\newblock \showarticletitle{I lead, you help but only with enough details:
  Understanding user experience of co-creation with artificial intelligence}.
  In \bibinfo{booktitle}{\emph{Proceedings of the 2018 CHI Conference on Human
  Factors in Computing Systems}}. \bibinfo{pages}{1--13}.
\newblock


\bibitem[\protect\citeauthoryear{O’Connor, Tsafnat, Thomas, Glasziou,
  Gilbert, and Hutton}{O’Connor et~al\mbox{.}}{2019}]%
        {o2019question}
\bibfield{author}{\bibinfo{person}{Annette~M O’Connor}, \bibinfo{person}{Guy
  Tsafnat}, \bibinfo{person}{James Thomas}, \bibinfo{person}{Paul Glasziou},
  \bibinfo{person}{Stephen~B Gilbert}, {and} \bibinfo{person}{Brian Hutton}.}
  \bibinfo{year}{2019}\natexlab{}.
\newblock \showarticletitle{A question of trust: can we build an evidence base
  to gain trust in systematic review automation technologies?}
\newblock \bibinfo{journal}{\emph{Systematic reviews}} \bibinfo{volume}{8},
  \bibinfo{number}{1} (\bibinfo{year}{2019}), \bibinfo{pages}{1--8}.
\newblock


\bibitem[\protect\citeauthoryear{Pangburn}{Pangburn}{2019}]%
        {pangburn2019schools}
\bibfield{author}{\bibinfo{person}{DJ Pangburn}.}
  \bibinfo{year}{2019}\natexlab{}.
\newblock \showarticletitle{Schools are using software to help pick who gets
  in. What could go wrong?}
\newblock \bibinfo{journal}{\emph{Fast Company}} (\bibinfo{date}{17 May}
  \bibinfo{year}{2019}).
\newblock
\urldef\tempurl%
\url{https://www.fastcompany.com/90342596/schools-are-quietly-
  turning-to-ai-to-help-pick-who-gets-in-what-could-go-wrong}
\showURL{%
Retrieved 05-October-2021 from \tempurl}


\bibitem[\protect\citeauthoryear{Panko}{Panko}{1998}]%
        {panko1998we}
\bibfield{author}{\bibinfo{person}{Raymond~R Panko}.}
  \bibinfo{year}{1998}\natexlab{}.
\newblock \showarticletitle{What we know about spreadsheet errors}.
\newblock \bibinfo{journal}{\emph{Journal of Organizational and End User
  Computing (JOEUC)}} \bibinfo{volume}{10}, \bibinfo{number}{2}
  (\bibinfo{year}{1998}), \bibinfo{pages}{15--21}.
\newblock


\bibitem[\protect\citeauthoryear{Parasuraman, Sheridan, and
  Wickens}{Parasuraman et~al\mbox{.}}{2008}]%
        {parasuraman2008situation}
\bibfield{author}{\bibinfo{person}{Raja Parasuraman}, \bibinfo{person}{Thomas~B
  Sheridan}, {and} \bibinfo{person}{Christopher~D Wickens}.}
  \bibinfo{year}{2008}\natexlab{}.
\newblock \showarticletitle{Situation awareness, mental workload, and trust in
  automation: Viable, empirically supported cognitive engineering constructs}.
\newblock \bibinfo{journal}{\emph{Journal of cognitive engineering and decision
  making}} \bibinfo{volume}{2}, \bibinfo{number}{2} (\bibinfo{year}{2008}),
  \bibinfo{pages}{140--160}.
\newblock


\bibitem[\protect\citeauthoryear{Puri, Kung, Janssen, Zhang, Domeniconi,
  Zolotov, Dolby, Chen, Choudhury, Decker, et~al\mbox{.}}{Puri
  et~al\mbox{.}}{2021}]%
        {puri2021project}
\bibfield{author}{\bibinfo{person}{Ruchir Puri}, \bibinfo{person}{David~S
  Kung}, \bibinfo{person}{Geert Janssen}, \bibinfo{person}{Wei Zhang},
  \bibinfo{person}{Giacomo Domeniconi}, \bibinfo{person}{Vladmir Zolotov},
  \bibinfo{person}{Julian Dolby}, \bibinfo{person}{Jie Chen},
  \bibinfo{person}{Mihir Choudhury}, \bibinfo{person}{Lindsey Decker},
  {et~al\mbox{.}}} \bibinfo{year}{2021}\natexlab{}.
\newblock \showarticletitle{Project CodeNet: A Large-Scale AI for Code Dataset
  for Learning a Diversity of Coding Tasks}.
\newblock \bibinfo{journal}{\emph{arXiv preprint arXiv:2105.12655}}
  (\bibinfo{year}{2021}).
\newblock


\bibitem[\protect\citeauthoryear{Radford, Wu, Child, Luan, Amodei, Sutskever,
  et~al\mbox{.}}{Radford et~al\mbox{.}}{2019}]%
        {radford2019language}
\bibfield{author}{\bibinfo{person}{Alec Radford}, \bibinfo{person}{Jeffrey Wu},
  \bibinfo{person}{Rewon Child}, \bibinfo{person}{David Luan},
  \bibinfo{person}{Dario Amodei}, \bibinfo{person}{Ilya Sutskever},
  {et~al\mbox{.}}} \bibinfo{year}{2019}\natexlab{}.
\newblock \showarticletitle{Language models are unsupervised multitask
  learners}.
\newblock \bibinfo{journal}{\emph{OpenAI blog}} \bibinfo{volume}{1},
  \bibinfo{number}{8} (\bibinfo{year}{2019}), \bibinfo{pages}{9}.
\newblock


\bibitem[\protect\citeauthoryear{Riedl}{Riedl}{2019}]%
        {riedl2019human}
\bibfield{author}{\bibinfo{person}{Mark~O Riedl}.}
  \bibinfo{year}{2019}\natexlab{}.
\newblock \showarticletitle{Human-centered artificial intelligence and machine
  learning}.
\newblock \bibinfo{journal}{\emph{Human Behavior and Emerging Technologies}}
  \bibinfo{volume}{1}, \bibinfo{number}{1} (\bibinfo{year}{2019}),
  \bibinfo{pages}{33--36}.
\newblock


\bibitem[\protect\citeauthoryear{R{\"o}nngren and Ayani}{R{\"o}nngren and
  Ayani}{1997}]%
        {ronngren1997comparative}
\bibfield{author}{\bibinfo{person}{Robert R{\"o}nngren} {and}
  \bibinfo{person}{Rassul Ayani}.} \bibinfo{year}{1997}\natexlab{}.
\newblock \showarticletitle{A comparative study of parallel and sequential
  priority queue algorithms}.
\newblock \bibinfo{journal}{\emph{ACM Transactions on Modeling and Computer
  Simulation (TOMACS)}} \bibinfo{volume}{7}, \bibinfo{number}{2}
  (\bibinfo{year}{1997}), \bibinfo{pages}{157--209}.
\newblock


\bibitem[\protect\citeauthoryear{Ross, Chen, Hang, Glassman, and
  Doshi-Velez}{Ross et~al\mbox{.}}{2021}]%
        {ross2021evaluating}
\bibfield{author}{\bibinfo{person}{Andrew Ross}, \bibinfo{person}{Nina Chen},
  \bibinfo{person}{Elisa~Zhao Hang}, \bibinfo{person}{Elena~L Glassman}, {and}
  \bibinfo{person}{Finale Doshi-Velez}.} \bibinfo{year}{2021}\natexlab{}.
\newblock \showarticletitle{Evaluating the Interpretability of Generative
  Models by Interactive Reconstruction}. In
  \bibinfo{booktitle}{\emph{Proceedings of the 2021 CHI Conference on Human
  Factors in Computing Systems}}. \bibinfo{pages}{1--15}.
\newblock


\bibitem[\protect\citeauthoryear{Rossi}{Rossi}{2018}]%
        {rossi2018building}
\bibfield{author}{\bibinfo{person}{Francesca Rossi}.}
  \bibinfo{year}{2018}\natexlab{}.
\newblock \showarticletitle{Building trust in artificial intelligence}.
\newblock \bibinfo{journal}{\emph{Journal of international affairs}}
  \bibinfo{volume}{72}, \bibinfo{number}{1} (\bibinfo{year}{2018}),
  \bibinfo{pages}{127--134}.
\newblock


\bibitem[\protect\citeauthoryear{Roziere, Lachaux, Chanussot, and
  Lample}{Roziere et~al\mbox{.}}{2020}]%
        {roziere2020unsupervised}
\bibfield{author}{\bibinfo{person}{Baptiste Roziere},
  \bibinfo{person}{Marie-Anne Lachaux}, \bibinfo{person}{Lowik Chanussot},
  {and} \bibinfo{person}{Guillaume Lample}.} \bibinfo{year}{2020}\natexlab{}.
\newblock \showarticletitle{Unsupervised Translation of Programming
  Languages.}. In \bibinfo{booktitle}{\emph{NeurIPS}}.
\newblock


\bibitem[\protect\citeauthoryear{Sankaran, Zhang, Gutierrez~Lopez, and
  V{\"a}{\"a}n{\"a}nen}{Sankaran et~al\mbox{.}}{2020}]%
        {sankaran2020respecting}
\bibfield{author}{\bibinfo{person}{Supraja Sankaran}, \bibinfo{person}{Chao
  Zhang}, \bibinfo{person}{Marisela Gutierrez~Lopez}, {and}
  \bibinfo{person}{Kaisa V{\"a}{\"a}n{\"a}nen}.}
  \bibinfo{year}{2020}\natexlab{}.
\newblock \showarticletitle{Respecting Human Autonomy through Human-Centered
  AI}. In \bibinfo{booktitle}{\emph{Proceedings of the 11th Nordic Conference
  on Human-Computer Interaction: Shaping Experiences, Shaping Society}}.
  \bibinfo{pages}{1--3}.
\newblock


\bibitem[\protect\citeauthoryear{Saxena, Badillo-Urquiola, Wisniewski, and
  Guha}{Saxena et~al\mbox{.}}{2021}]%
        {saxena2021framework}
\bibfield{author}{\bibinfo{person}{Devansh Saxena}, \bibinfo{person}{Karla
  Badillo-Urquiola}, \bibinfo{person}{Pamela Wisniewski}, {and}
  \bibinfo{person}{Shion Guha}.} \bibinfo{year}{2021}\natexlab{}.
\newblock \showarticletitle{A Framework of High-Stakes Algorithmic
  Decision-Making for the Public Sector Developed through a Case Study of
  Child-Welfare}.
\newblock \bibinfo{journal}{\emph{arXiv preprint arXiv:2107.03487}}
  (\bibinfo{year}{2021}).
\newblock


\bibitem[\protect\citeauthoryear{Saxena, Badillo-Urquiola, Wisniewski, and
  Guha}{Saxena et~al\mbox{.}}{2020}]%
        {saxena2020human}
\bibfield{author}{\bibinfo{person}{Devansh Saxena}, \bibinfo{person}{Karla
  Badillo-Urquiola}, \bibinfo{person}{Pamela~J Wisniewski}, {and}
  \bibinfo{person}{Shion Guha}.} \bibinfo{year}{2020}\natexlab{}.
\newblock \showarticletitle{A Human-Centered Review of Algorithms used within
  the US Child Welfare System}. In \bibinfo{booktitle}{\emph{Proceedings of the
  2020 CHI Conference on Human Factors in Computing Systems}}.
  \bibinfo{pages}{1--15}.
\newblock


\bibitem[\protect\citeauthoryear{Schneider, Ghellal, Love, and
  Gerlicher}{Schneider et~al\mbox{.}}{2021}]%
        {schneider2021increasing}
\bibfield{author}{\bibinfo{person}{Tobias Schneider}, \bibinfo{person}{Sabiha
  Ghellal}, \bibinfo{person}{Steve Love}, {and} \bibinfo{person}{Ansgar~RS
  Gerlicher}.} \bibinfo{year}{2021}\natexlab{}.
\newblock \showarticletitle{Increasing the User Experience in Autonomous
  Driving through different Feedback Modalities}. In
  \bibinfo{booktitle}{\emph{26th International Conference on Intelligent User
  Interfaces}}. \bibinfo{pages}{7--10}.
\newblock


\bibitem[\protect\citeauthoryear{Schuler and Namioka}{Schuler and
  Namioka}{1993}]%
        {schuler1993participatory}
\bibfield{author}{\bibinfo{person}{Douglas Schuler} {and} \bibinfo{person}{Aki
  Namioka}.} \bibinfo{year}{1993}\natexlab{}.
\newblock \bibinfo{booktitle}{\emph{Participatory design: Principles and
  practices}}.
\newblock \bibinfo{publisher}{CRC Press}.
\newblock


\bibitem[\protect\citeauthoryear{Schwabe and Castellacci}{Schwabe and
  Castellacci}{2020}]%
        {schwabe2020automation}
\bibfield{author}{\bibinfo{person}{Henrik Schwabe} {and}
  \bibinfo{person}{Fulvio Castellacci}.} \bibinfo{year}{2020}\natexlab{}.
\newblock \showarticletitle{Automation, workers’ skills and job
  satisfaction}.
\newblock \bibinfo{journal}{\emph{Plos one}} \bibinfo{volume}{15},
  \bibinfo{number}{11} (\bibinfo{year}{2020}), \bibinfo{pages}{e0242929}.
\newblock


\bibitem[\protect\citeauthoryear{Seeber, Bittner, Briggs, De~Vreede, De~Vreede,
  Elkins, Maier, Merz, Oeste-Rei{\ss}, Randrup, et~al\mbox{.}}{Seeber
  et~al\mbox{.}}{2020}]%
        {seeber2020machines}
\bibfield{author}{\bibinfo{person}{Isabella Seeber}, \bibinfo{person}{Eva
  Bittner}, \bibinfo{person}{Robert~O Briggs}, \bibinfo{person}{Triparna
  De~Vreede}, \bibinfo{person}{Gert-Jan De~Vreede}, \bibinfo{person}{Aaron
  Elkins}, \bibinfo{person}{Ronald Maier}, \bibinfo{person}{Alexander~B Merz},
  \bibinfo{person}{Sarah Oeste-Rei{\ss}}, \bibinfo{person}{Nils Randrup},
  {et~al\mbox{.}}} \bibinfo{year}{2020}\natexlab{}.
\newblock \showarticletitle{Machines as teammates: A research agenda on AI in
  team collaboration}.
\newblock \bibinfo{journal}{\emph{Information \& management}}
  \bibinfo{volume}{57}, \bibinfo{number}{2} (\bibinfo{year}{2020}),
  \bibinfo{pages}{103174}.
\newblock


\bibitem[\protect\citeauthoryear{Shneiderman}{Shneiderman}{2020a}]%
        {shneiderman2020bridging}
\bibfield{author}{\bibinfo{person}{Ben Shneiderman}.}
  \bibinfo{year}{2020}\natexlab{a}.
\newblock \showarticletitle{Bridging the gap between ethics and practice:
  Guidelines for reliable, safe, and trustworthy Human-Centered AI systems}.
\newblock \bibinfo{journal}{\emph{ACM Transactions on Interactive Intelligent
  Systems (TiiS)}} \bibinfo{volume}{10}, \bibinfo{number}{4}
  (\bibinfo{year}{2020}), \bibinfo{pages}{1--31}.
\newblock


\bibitem[\protect\citeauthoryear{Shneiderman}{Shneiderman}{2020b}]%
        {shneiderman2020human}
\bibfield{author}{\bibinfo{person}{Ben Shneiderman}.}
  \bibinfo{year}{2020}\natexlab{b}.
\newblock \showarticletitle{Human-centered artificial intelligence: Three fresh
  ideas}.
\newblock \bibinfo{journal}{\emph{AIS Transactions on Human-Computer
  Interaction}} \bibinfo{volume}{12}, \bibinfo{number}{3}
  (\bibinfo{year}{2020}), \bibinfo{pages}{109--124}.
\newblock


\bibitem[\protect\citeauthoryear{Shneiderman}{Shneiderman}{2021}]%
        {shneiderman2021tutorial}
\bibfield{author}{\bibinfo{person}{Ben Shneiderman}.}
  \bibinfo{year}{2021}\natexlab{}.
\newblock \showarticletitle{Tutorial: Human-Centered AI: Reliable, Safe and
  Trustworthy}. In \bibinfo{booktitle}{\emph{26th International Conference on
  Intelligent User Interfaces}}. \bibinfo{pages}{7--8}.
\newblock


\bibitem[\protect\citeauthoryear{Spinuzzi}{Spinuzzi}{2005}]%
        {spinuzzi2005methodology}
\bibfield{author}{\bibinfo{person}{Clay Spinuzzi}.}
  \bibinfo{year}{2005}\natexlab{}.
\newblock \showarticletitle{The methodology of participatory design}.
\newblock \bibinfo{journal}{\emph{Technical communication}}
  \bibinfo{volume}{52}, \bibinfo{number}{2} (\bibinfo{year}{2005}),
  \bibinfo{pages}{163--174}.
\newblock


\bibitem[\protect\citeauthoryear{Szymanski, Millecamp, and Verbert}{Szymanski
  et~al\mbox{.}}{2021}]%
        {szymanski2021visual}
\bibfield{author}{\bibinfo{person}{Maxwell Szymanski}, \bibinfo{person}{Martijn
  Millecamp}, {and} \bibinfo{person}{Katrien Verbert}.}
  \bibinfo{year}{2021}\natexlab{}.
\newblock \showarticletitle{Visual, textual or hybrid: the effect of user
  expertise on different explanations}. In \bibinfo{booktitle}{\emph{26th
  International Conference on Intelligent User Interfaces}}.
  \bibinfo{pages}{109--119}.
\newblock


\bibitem[\protect\citeauthoryear{Tufano, Drain, Svyatkovskiy, Deng, and
  Sundaresan}{Tufano et~al\mbox{.}}{2020}]%
        {tufano2020unit}
\bibfield{author}{\bibinfo{person}{Michele Tufano}, \bibinfo{person}{Dawn
  Drain}, \bibinfo{person}{Alexey Svyatkovskiy}, \bibinfo{person}{Shao~Kun
  Deng}, {and} \bibinfo{person}{Neel Sundaresan}.}
  \bibinfo{year}{2020}\natexlab{}.
\newblock \showarticletitle{Unit Test Case Generation with Transformers}.
\newblock \bibinfo{journal}{\emph{arXiv preprint arXiv:2009.05617}}
  (\bibinfo{year}{2020}).
\newblock


\bibitem[\protect\citeauthoryear{Wang, Wang, Drozdal, Muller, Park, Weisz, Liu,
  Wu, and Dugan}{Wang et~al\mbox{.}}{2021}]%
        {wang2021themisto}
\bibfield{author}{\bibinfo{person}{April~Yi Wang}, \bibinfo{person}{Dakuo
  Wang}, \bibinfo{person}{Jaimie Drozdal}, \bibinfo{person}{Michael Muller},
  \bibinfo{person}{Soya Park}, \bibinfo{person}{Justin~D Weisz},
  \bibinfo{person}{Xuye Liu}, \bibinfo{person}{Lingfei Wu}, {and}
  \bibinfo{person}{Casey Dugan}.} \bibinfo{year}{2021}\natexlab{}.
\newblock \showarticletitle{Themisto: Towards Automated Documentation
  Generation in Computational Notebooks}.
\newblock \bibinfo{journal}{\emph{arXiv preprint arXiv:2102.12592}}
  (\bibinfo{year}{2021}).
\newblock


\bibitem[\protect\citeauthoryear{Wang, Ram, Weidele, Liu, Muller, Weisz,
  Valente, Chaudhary, Torres, Samulowitz, et~al\mbox{.}}{Wang
  et~al\mbox{.}}{2020}]%
        {wang2020autoai}
\bibfield{author}{\bibinfo{person}{Dakuo Wang}, \bibinfo{person}{Parikshit
  Ram}, \bibinfo{person}{Daniel Karl~I Weidele}, \bibinfo{person}{Sijia Liu},
  \bibinfo{person}{Michael Muller}, \bibinfo{person}{Justin~D Weisz},
  \bibinfo{person}{Abel Valente}, \bibinfo{person}{Arunima Chaudhary},
  \bibinfo{person}{Dustin Torres}, \bibinfo{person}{Horst Samulowitz},
  {et~al\mbox{.}}} \bibinfo{year}{2020}\natexlab{}.
\newblock \showarticletitle{Autoai: Automating the end-to-end ai lifecycle with
  humans-in-the-loop}. In \bibinfo{booktitle}{\emph{Proceedings of the 25th
  International Conference on Intelligent User Interfaces Companion}}.
  \bibinfo{pages}{77--78}.
\newblock


\bibitem[\protect\citeauthoryear{Wang, Weisz, Muller, Ram, Geyer, Dugan,
  Tausczik, Samulowitz, and Gray}{Wang et~al\mbox{.}}{2019}]%
        {wang2019human}
\bibfield{author}{\bibinfo{person}{Dakuo Wang}, \bibinfo{person}{Justin~D
  Weisz}, \bibinfo{person}{Michael Muller}, \bibinfo{person}{Parikshit Ram},
  \bibinfo{person}{Werner Geyer}, \bibinfo{person}{Casey Dugan},
  \bibinfo{person}{Yla Tausczik}, \bibinfo{person}{Horst Samulowitz}, {and}
  \bibinfo{person}{Alexander Gray}.} \bibinfo{year}{2019}\natexlab{}.
\newblock \showarticletitle{Human-ai collaboration in data science: Exploring
  data scientists' perceptions of automated ai}.
\newblock \bibinfo{journal}{\emph{Proceedings of the ACM on Human-Computer
  Interaction}} \bibinfo{volume}{3}, \bibinfo{number}{CSCW}
  (\bibinfo{year}{2019}), \bibinfo{pages}{1--24}.
\newblock


\bibitem[\protect\citeauthoryear{Wang and Yin}{Wang and Yin}{2021}]%
        {wang2021explanations}
\bibfield{author}{\bibinfo{person}{Xinru Wang} {and} \bibinfo{person}{Ming
  Yin}.} \bibinfo{year}{2021}\natexlab{}.
\newblock \showarticletitle{Are Explanations Helpful? A Comparative Study of
  the Effects of Explanations in AI-Assisted Decision-Making}. In
  \bibinfo{booktitle}{\emph{26th International Conference on Intelligent User
  Interfaces}}. \bibinfo{pages}{318--328}.
\newblock


\bibitem[\protect\citeauthoryear{Warner and Guo}{Warner and Guo}{2017}]%
        {warner2017codepilot}
\bibfield{author}{\bibinfo{person}{Jeremy Warner} {and}
  \bibinfo{person}{Philip~J Guo}.} \bibinfo{year}{2017}\natexlab{}.
\newblock \showarticletitle{Codepilot: Scaffolding end-to-end collaborative
  software development for novice programmers}. In
  \bibinfo{booktitle}{\emph{Proceedings of the 2017 CHI Conference on Human
  Factors in Computing Systems}}. \bibinfo{pages}{1136--1141}.
\newblock


\bibitem[\protect\citeauthoryear{Weber, Hu{\ss}mann, Han, Matthes, and
  Liu}{Weber et~al\mbox{.}}{2020}]%
        {weber2020draw}
\bibfield{author}{\bibinfo{person}{Thomas Weber}, \bibinfo{person}{Heinrich
  Hu{\ss}mann}, \bibinfo{person}{Zhiwei Han}, \bibinfo{person}{Stefan Matthes},
  {and} \bibinfo{person}{Yuanting Liu}.} \bibinfo{year}{2020}\natexlab{}.
\newblock \showarticletitle{Draw with me: Human-in-the-loop for image
  restoration}. In \bibinfo{booktitle}{\emph{Proceedings of the 25th
  International Conference on Intelligent User Interfaces}}.
  \bibinfo{pages}{243--253}.
\newblock


\bibitem[\protect\citeauthoryear{Weisz, Muller, Houde, Richards, Ross,
  Martinez, Agarwal, and Talamadupula}{Weisz et~al\mbox{.}}{2021}]%
        {weisz2021perfection}
\bibfield{author}{\bibinfo{person}{Justin~D Weisz}, \bibinfo{person}{Michael
  Muller}, \bibinfo{person}{Stephanie Houde}, \bibinfo{person}{John Richards},
  \bibinfo{person}{Steven~I Ross}, \bibinfo{person}{Fernando Martinez},
  \bibinfo{person}{Mayank Agarwal}, {and} \bibinfo{person}{Kartik
  Talamadupula}.} \bibinfo{year}{2021}\natexlab{}.
\newblock \showarticletitle{Perfection Not Required? Human-AI Partnerships in
  Code Translation}. In \bibinfo{booktitle}{\emph{26th International Conference
  on Intelligent User Interfaces}}. \bibinfo{pages}{402--412}.
\newblock


\bibitem[\protect\citeauthoryear{Widder, Dabbish, Herbsleb, Holloway, and
  Davidoff}{Widder et~al\mbox{.}}{2021}]%
        {widder2021trust}
\bibfield{author}{\bibinfo{person}{David~Gray Widder}, \bibinfo{person}{Laura
  Dabbish}, \bibinfo{person}{James~D Herbsleb}, \bibinfo{person}{Alexandra
  Holloway}, {and} \bibinfo{person}{Scott Davidoff}.}
  \bibinfo{year}{2021}\natexlab{}.
\newblock \showarticletitle{Trust in Collaborative Automation in High Stakes
  Software Engineering Work: A Case Study at NASA}. In
  \bibinfo{booktitle}{\emph{Proceedings of the 2021 CHI Conference on Human
  Factors in Computing Systems}}. \bibinfo{pages}{1--13}.
\newblock


\bibitem[\protect\citeauthoryear{Wiehr, Cakar, Daiber, and Kr{\"u}ger}{Wiehr
  et~al\mbox{.}}{2021}]%
        {wiehr2021effect}
\bibfield{author}{\bibinfo{person}{Frederik Wiehr}, \bibinfo{person}{Baris
  Cakar}, \bibinfo{person}{Florian Daiber}, {and} \bibinfo{person}{Antonio
  Kr{\"u}ger}.} \bibinfo{year}{2021}\natexlab{}.
\newblock \showarticletitle{The Effect of Surrounding Scenery Complexity on the
  Transfer of Control Time in Highly Automated Driving}. In
  \bibinfo{booktitle}{\emph{26th International Conference on Intelligent User
  Interfaces}}. \bibinfo{pages}{92--97}.
\newblock


\bibitem[\protect\citeauthoryear{Wolf}{Wolf}{2020}]%
        {wolf2020democratizing}
\bibfield{author}{\bibinfo{person}{Christine~T Wolf}.}
  \bibinfo{year}{2020}\natexlab{}.
\newblock \showarticletitle{Democratizing AI? experience and accessibility in
  the age of artificial intelligence}.
\newblock \bibinfo{journal}{\emph{XRDS: Crossroads, The ACM Magazine for
  Students}} \bibinfo{volume}{26}, \bibinfo{number}{4} (\bibinfo{year}{2020}),
  \bibinfo{pages}{12--15}.
\newblock


\bibitem[\protect\citeauthoryear{Xu, Vasilescu, and Neubig}{Xu
  et~al\mbox{.}}{2021}]%
        {xu2021ide}
\bibfield{author}{\bibinfo{person}{Frank~F Xu}, \bibinfo{person}{Bogdan
  Vasilescu}, {and} \bibinfo{person}{Graham Neubig}.}
  \bibinfo{year}{2021}\natexlab{}.
\newblock \showarticletitle{In-IDE Code Generation from Natural Language:
  Promise and Challenges}.
\newblock \bibinfo{journal}{\emph{arXiv preprint arXiv:2101.11149}}
  (\bibinfo{year}{2021}).
\newblock


\bibitem[\protect\citeauthoryear{Xu}{Xu}{2019}]%
        {xu2019toward}
\bibfield{author}{\bibinfo{person}{Wei Xu}.} \bibinfo{year}{2019}\natexlab{}.
\newblock \showarticletitle{Toward human-centered AI: a perspective from
  human-computer interaction}.
\newblock \bibinfo{journal}{\emph{Interactions}} \bibinfo{volume}{26},
  \bibinfo{number}{4} (\bibinfo{year}{2019}), \bibinfo{pages}{42--46}.
\newblock


\end{thebibliography}

\clearpage
\appendix

\section{Data Structure Implementations}
\label{appendix:data-structures}

\subsection{Trie}
Our Java implementation of the Trie consisted of 170 lines (93 SLOC). It consisted of two classes, \texttt{TrieNode} and \texttt{Trie}. The \texttt{TrieNode} class was a small helper class for storing data in the \texttt{Trie} and consisted of four one-line methods. The \texttt{Trie} class defined additional methods for inserting strings and testing membership, deleting a string, enumerating all strings, and merging with another \texttt{Trie}. Two functions -- \texttt{delete()} and \texttt{enumerate()} -- were implemented as a pair (a primary function and a recursive helper), taking advantage of Java's support for method overloading, in which two methods share the same name but have different signatures. However, as Python does not allow duplicate method names, part of the translation process involved figuring out an appropriate way to solve this issue. Two options were possible: 1) defining a single method that used default arguments to pass data recursively (e.g. \texttt{def delete(self, word, current=None, index=0)}), or 2) giving the recursive helper method a different name (e.g. \texttt{\_delete()}).

We intentionally included this design option to force participants to think deeply about how to translate code when a feature of one language -- Java's ability to have duplicate method names -- isn't present in the other. Because of the alternate solutions to this problem, participants' translations of the \texttt{Trie} differed in the number of methods defined: 10 when both \texttt{delete()} and \texttt{enumerate()} were implemented with a single method, 11 when one function was implemented with a single method and the other with two, or 12 when both functions were implemented with two methods. This detail is important for our computation of the proportion of correctly-implemented methods in Section~\ref{sec:code-measures}, and thus, we counted both the number of methods implemented in the \texttt{Trie} as well as the number implemented correctly.

\subsection{Priority Queue}
Our Java implementation of the Priority Queue consisted of 157 lines (83 SLOC). It consisted of a single \texttt{PriorityQueue} class with nine methods for constructing a queue, inserting and removing elements, inserting a collection of elements, peeking the head of the queue, determining if the queue is empty, retrieving the size of the queue, enumerating all elements, and an internal method for restoring the heap property.

\section{Producing Code Translations with TransCoder}
\label{appendix:ai-produced-translations}

To create the AI-produced translations, we employed the TransCoder model~\cite{roziere2020unsupervised}. Although this model performs fairly well on standalone methods, it was unable to handle the translation of whole classes. Therefore, we translated the Java classes on a method-by-method basis, with a small degree of manual cleanup of method signatures and instance data access (e.g. adding \texttt{self} to instance member references).

In order to alter the quality of the translations, we introduced a new beam offset parameter to the model. This parameter allowed us to guide the beam search by adjusting which tokens were taken during each phase of the search. TransCoder's default behavior is to consider the top-k most likely tokens, resulting in our set of better-quality translations. By adjusting the beam offset, we forced the model to discard the top-k most likely tokens, and consider other, less-likely tokens. To produce the set of worse-quality translations, we used a beam offset parameter of 10.

\subsection{Spurious Code Examples}
\label{appendix:spurious-code}

In some instances, NMT models such as TransCoder produce code that is \emph{spurious}: unrelated or irrelevant to the input provided to the model. Figure~\ref{fig:off-the-rails} shows two such examples. In example (a), the model produced code that references an undefined class named \texttt{Value}. It also includes code that constructs string representations of the elements in a list -- \texttt{"".join(...)} -- for which there was no equivalent in the source Java. In example (b), multiple methods of the \texttt{os} package were invoked, even though their use was unnecessary and there was no equivalent use in the source Java.

Other instances of spurious code included by the model included unnecessary \texttt{import} statements (e.g. \texttt{import os}), unnecessary \texttt{return} statements in methods that did not return a value, extra arguments present in method calls, duplicate definitions of class initializers, and calls to undefined methods.

\begin{figure}
    \small
    \setlength{\belowcaptionskip}{1.0\baselineskip}
    \begin{subfigure}[t]{1.0\linewidth}
    \begin{minted}[mathescape,
                       numbersep=5pt,
                       frame=lines,
                       framesep=2mm,
                       breaklines]{python}
return Value(
    "".join(entry[0] for entry in entries),		
    root="".join(entry[1] for entry in entries),		
    verbose=True,		
)
    \end{minted}
    \caption{Example Python code from a worse-quality translation of the Priority Queue that references an undefined class (\texttt{Value}) and includes code not relevant to the Java implementation.}
    \label{degenerate-output-pq}
  \end{subfigure}
  \par
  \begin{subfigure}[t]{1.0\linewidth}
    \centering
    \begin{minted}[mathescape,
                      numbersep=5pt,
                      frame=lines,
                      framesep=2mm,
                      breaklines]{python}
enumerate(
    [		
        root		
        for root, dirs, files in os.walk(". ")		
        if not os.path.exists(os.path.join(root, "__init__.py"))		
    ]
)
        \end{minted}
    \caption{Example Python code from a better-quality translation of the Trie that includes code not relevant to the Java implementation.}
    \label{degenerate-output-trie}
  \end{subfigure}
  \caption{Example cases of nonsensical outputs from the code translation model. In both cases, the code references classes and/or methods that were not defined or were not relevant for the task at hand. Examples are from (a) a worse-quality translation of the Priority Queue, and (b) a better-quality translation of the Trie.}
  \label{fig:off-the-rails}
\end{figure}

\section{Reproducibility Package}
\label{appendix:reproducibility-package}

\subsection{Code}
\label{appendix:reproducibility-code}

We have published our Java data structure implementations and AI-produced code translations in Github at  \url{https://github.com/jweisz/iui22-code-translation}. We have also included codebooks showing our error analysis of the AI translations to demonstrate how we coded the different kinds of programming errors detailed in Table~\ref{tab:error-taxonomy}.

Our motivation in publishing this code is to enable others to reproduce of our work. However, we would instead challenge the community to extend our work, using these materials as a reference. The specific source examples in the repository are limited to only one kind of code-related task (Java to Python translation) for which generative code models can provide aid. There are many other kinds of tasks for which generative code models can provide support, such as natural language to code, code documentation, code autocomplete, test case generation, bug repair, and others. We encourage further research into those use cases, especially focused on how intelligent user interfaces can help users achieve successful outcomes when working in the presence of imperfect model output.

\subsection{Participant Instructions}
\label{appendix:participant-instructions}

The scripts below were used to introduce the study and each of the code translation tasks to participants. Notes for the experimenter are denoted in \textbf{bold text}, and places where the script depends on the participant's condition are indicated with \emph{italicized text.}

\subsubsection{Initial Instructions}
Thanks for participating in our study of AI-supported coding. Before we begin, I'd like to explain what this study is about and what you will be expected to do as a participant.

In this study, we are evaluating whether a specific kind of AI support is helpful in completing a coding task. We will ask you to perform two coding tasks, one with the AI support and one without it, so that we can ask you to contrast your experiences between the two. These tasks will be done in a random order.

In general, these tasks involve translating a data structure from Java to Python, and you'll have 30 minutes to do this. However, there's an important point that I must emphasize: there is more work to do in these tasks than what can be accomplished within 30 minutes. The reason for this is because we're not evaluating you or your software engineering skills, we're really trying to understand whether or not the AI support is actually helpful.

This point is so important that I need to say it again: we are not evaluating you or your programming skills in this study, we are evaluating whether the AI support is actually helpful or not.

After completing each task, I'll ask you to fill out a short survey about your work on the task, and after you finish both tasks, I'll ask you to fill out another survey about the entire experience.

There are a few more quick things I need to mention about this study before we begin:

\begin{itemize}
    \item Any data we publish on this study will be anonymized or aggregated across participants. This means, for example, if we quote anything that you say during the study, it will be anonymous and your name won't be attached to it.

    \item There's no penalty or cost to you if you decide to drop out right now. You can also stop participating in the study at any time, just by telling me you'd like to stop.
\end{itemize}

Do you have any questions about any of this? If you are OK with participating in this study, I'll need your verbal consent. Would you like to continue to participate?

\textbf{Continue only if verbal consent is provided.}

Great! Let's get started. Your first task will be converting an implementation of the \emph{\{trie, priority queue\}} data structure from Java to Python, and you'll do this one \emph{\{with, without\}} the AI support. Your second task will be converting an implementation of the \emph{\{trie, priority queue\}} data structure from Java to Python, and you'll do that one \emph{\{with, without\}} the AI support. We'll also take a break between the two tasks since this session will take about 2 hours.

First, let me tell you a bit more about this code translation task. The task is to produce a complete and working Python translation within 30 minutes. By complete, I mean that all of the functionality and documentation from the Java code should be translated. And by working, I mean that the Python code should be tested for correctness.

As I mentioned earlier, there's more work to do to accomplish an absolutely perfect translation than what can be done within 30 minutes. So really, I'm asking you to just do as much as you can in the time that you have. But, there may be tradeoffs in how you spend your time, such as whether you choose to translate all of the functionality first, or to translate a portion of it while testing it to make sure it works. How you navigate this tradeoff is completely up to you.

There are a few other things I should mention about this task:

\begin{itemize}
    \item You'll be using an online version of the VSCode editor in this study, which has a few minor differences from the desktop app. In general, all of the menu commands and keystrokes are available and should just work, but please do let me know if you encounter any difficulties.
    \item I won't be able to answer any technical questions about the code or what it's supposed to do, but I can tell you that the Java implementation is correct.
    \item When you make the translation, please try to preserve the naming of class and method names, but we are less concerned with how you name variables.
    \item You can also use whatever resources you feel would be helpful, such as searching the web or Stack Overflow, or running any of the code, since we've installed Java and Python in the editor.
    \item I'll give you a 5 minute warning when time is almost up so you know to start wrapping up your work.
\end{itemize}

We also ask that as you work on the tasks, you think out loud and tell us about what you're thinking and what you're doing. This really helps us understand how you approach the task and what you're thinking about as you do it. If you're quiet, I'll remind you to keep thinking out loud.

Do you have any questions before we get started? OK, let's begin!

\subsubsection{Task Instructions: AI Support}

In this task, you'll translate the \emph{\{trie, priority queue\}} data structure from Java to Python with the help of an AI code translation system. This system is able to automatically translate code from Java to Python, but it's not a perfect system, so the translated code may contain errors in it. For this task, please put your solution in the \emph{\{Trie.py, PriorityQueue.py\}} file.

\emph{If one translation:} We used this system ahead of time to create a Python translation for you, which you'll see when you open the editor. How you use this translation to help you complete the task is entirely up to you.

\emph{If five translations:} We used this system ahead of time to create five separate Python translations for you, which you'll see when you open the editor. How you use these translations to help you complete the task is entirely up to you.

\subsubsection{Task Instructions: No AI Support}
In this task, you'll translate the \emph{\{trie, priority queue\}} data structure from Java to Python, and you'll do this one without the AI support. For this task, please put your solution in the \emph{\{Trie.py, PriorityQueue.py\}} file.

\subsection{Surveys}

\subsubsection{Pre-Screen Survey}
\label{appendix:pre-screen-survey}

The questions in Table~\ref{tab:pre-screen} were used to screen potential participants for their experience with Java and Python, their familiarity with various data structures, and their familiarity with various code editing environments.

\begin{table*}[htp]
    \centering
    \small
    \begin{tabularx}{\textwidth}{XX}
        \toprule
        \textbf{Java} \\
        \midrule
        1. To what extent are you familiar with Java?
            & - I am not familiar with Java \\
            & - I have < 1 year of experience with Java \\
            & - I have 1-3 years experience with Java \\
            & - I have 3+ years of experience with Java \\
        2. How recently have you written Java code?
            & - Within the past month \\
            & - Within the past year \\
            & - Within the past 5 years \\
            & - Have not written Java code within the past 5 years \\
        \midrule
        \textbf{Python} \\
        \midrule
        3. To what extent are you familiar with Python? 
            & - I am not familiar with Python \\
            & - I have < 1 year of experience with Python \\
            & - I have 1-3 years experience with Python \\
            & - I have 3+ years of experience with Python \\
        4. How recently have you written Python code?
            & - Within the past month \\
            & - Within the past year \\
            & - Within the past 5 years \\
            & - Have not written Python code within the past 5 years \\
        \midrule
        \textbf{Data Structures} \\
        \midrule
        5. Please indicate the extent to which you are familiar with the following data structures.
        \begin{enumerate}[leftmargin=*, label=\alph*.]
            \item Binary Search Tree
            \item Splay Tree
            \item Trie
            \item Red/Black Tree
            \item B-Tree
            \item Heap / Priority Queue
            \item Union Find
        \end{enumerate}
        &
        \begin{itemize}[leftmargin=*, label={-}]
            \item Not familiar with it
            \item Familiar with it but have not used it within the past year
            \item Familiar with it and have used an implementation of it within the past year
            \item Familiar with it and have written my own implementation of it within the past year
        \end{itemize}
        \\
        \midrule
        \textbf{Programming Tools} \\
        \midrule
        6. Please indicate your familiarity with the following programming tools.
        \begin{enumerate}[leftmargin=*, label=\alph*.]
            \item Git
            \item PyCharm
            \item Eclipse
            \item VSCode
            \item Xcode
        \end{enumerate}
        &
        \begin{itemize}[leftmargin=*, label={-}]
            \item Not familiar with it
            \item Familiar with it but have not used it within the past year
            \item Familiar with it and have used it within the past year
            \item Familiar with it and use it on a regular basis
        \end{itemize}
        \\
        \bottomrule
    \end{tabularx}
    \caption{Pre-screening Questions}
    \label{tab:pre-screen}
\end{table*}

\subsubsection{Post-Task Survey}
\label{appendix:post-task-survey}

The questions in Table~\ref{tab:post-task-survey} were asked immediately after each code translation task. This survey took approximately 5-10 minutes to complete.

\begin{table*}[htp]
    \centering
    \small
    \begin{tabularx}{\textwidth}{Xp{7.8cm}}
        \toprule
        \textbf{Effort (NASA TLX~\cite{hart1988development})} \\
        \midrule
        1. How mentally demanding was the task?
            & {\small (Very Low) 1 2 3 4 5 6 7 8 9 10 11 12 13 14 15 16 17 18 19 20 (Very High)} \\
        2. How hurried or rushed was the pace of the task? 
            & {\small (Very Low) 1 2 3 4 5 6 7 8 9 10 11 12 13 14 15 16 17 18 19 20 (Very High)} \\
        3. How successful were you in accomplishing what you were asked to do?
            & {\small (Very Low) 1 2 3 4 5 6 7 8 9 10 11 12 13 14 15 16 17 18 19 20 (Very High)} \\
        4. How hard did you have to work to accomplish your level of performance?
            & {\small (Very Low) 1 2 3 4 5 6 7 8 9 10 11 12 13 14 15 16 17 18 19 20 (Very High)} \\
        5. How insecure, discouraged, irritated, stressed, and annoyed were you? 
            & {\small (Very Low) 1 2 3 4 5 6 7 8 9 10 11 12 13 14 15 16 17 18 19 20 (Very High)} \\
        \midrule
        \textbf{Work Process} \\
        \midrule
        1. Please describe your overall approach for how you worked on this task and how you allocated your time across different activities.
            & \emph{Open-ended} \\
        \midrule
        \textbf{AI Support} \emph{(only included in the AI condition}) \\
        \midrule
        1. How would you characterize the quality of the provided translation(s)?
            & Very good, Good, Acceptable, Poor, Very poor \\
        \midrule
        2. To what extent did you feel the provided translation(s)...
        \begin{enumerate}[leftmargin=*, label=\alph*.]
            \item were useful?
            \item had errors?
            \item helped you complete the task?
            \item taught you something new about Python?
        \end{enumerate}
            & Not at all, A little, Somewhat, A great deal \\
        \midrule
        3. Did you feel that the provided translation(s) made your work in this task difficult vs. easy?
            & (difficult) -3 -2 -1 0 +1 +2 +3 (easy) \\
        \midrule
        4. Did you feel that the provided translation(s) made your work in this task slow vs. fast?
            & (slow) -3 -2 -1 0 +1 +2 +3 (fast) \\
        \midrule
        5. Did you feel that the provided translation(s) made your work in this task of the worst vs. of the best quality?
            & (of the worst quality) -3 -2 -1 0 +1 +2 +3 (of the best quality) \\
        \midrule
        6. What did you find most useful about the provided translation(s)?
            & \emph{Open-ended} \\
        \midrule
        7. What did you find least useful about the provided translation(s)?
            & \emph{Open-ended} \\
        \midrule
        8. Were enough translations provided for you?
            & - I would have liked to have seen more translations \\
            & - I was provided with enough translations \\
            & - I was provided with too many translations \\
        \midrule
        9. What additional information could we have provided alongside the translation(s) to make them more useful to you? What additional information would you expect as part of an explanation of how the translation was made?
            & \emph{Open-ended} \\
        \midrule
        \textbf{Other Feedback} \\
        \midrule
        1. Is there any other feedback you would like to share with us about your experience on this task?
            & \emph{Open-ended} \\
        \bottomrule
    \end{tabularx}
    \caption{Post-task Survey}
    \label{tab:post-task-survey}
\end{table*}

\subsubsection{Post-Study Survey}
\label{appendix:post-study-survey}

The questions in Table~\ref{tab:post-study} were asked immediately after a participant completed the second post task survey. This survey took approximately 5-10 minutes to complete.

\begin{table*}[htp]
    \centering
    \small
    \begin{tabularx}{\textwidth}{XX}
        \toprule
        \textbf{Post-study Questions} \\
        \midrule
        1. How would you compare your experience working on the translation task without AI support vs. with AI support?
        & \emph{Open-ended} \\
        \midrule
        3. To which gender identity do you most identify?
            & - Male \\
            & - Female \\
            & - Transgender Male \\
            & - Transgender Female \\
            & - Gender Variant / Non-conforming \\
            & - Other (Write-in) \\
            & - Prefer not to answer \\
        \midrule
        4. Do you consider yourself primarily a...
            & - Data Scientist \\
            & - Designer \\
            & - Manager \\
            & - Researcher / Scientist \\
            & - Software Architect \\
            & - Software Engineer \\
            & - Other (Write-in) \\
        \midrule
        \multirow[t]{4}{\linewidth}{5. Have you used AI or machine learning algorithms as part of your work?}
            & - No \\
            & - Yes, within the past month \\
            & - Yes, within the past year \\
            & - Yes, but over one year ago \\
        \midrule
        6. Please explain your use of AI or machine learning algorithms.
            & \emph{Open-ended} \\
        \midrule
        \multirow[t]{4}{\linewidth}{7. Have you performed code translation tasks in your work?}
            & - No \\
            & - Yes, within the past month \\
            & - Yes, within the past year \\
            & - Yes, but over one year ago \\
        \midrule
        8. Please explain the kind of code translation work you have performed.
            & \emph{Open-ended} \\
        \bottomrule
    \end{tabularx}
    \caption{Post-study Survey}
    \label{tab:post-study}
\end{table*}

\end{document}